\documentclass[12pt,showpacs,preprintnumbers,amsmath,amssymb]{revtex4}
\usepackage{graphicx}
\begin{document}

\newcommand{\pderiv}[2]{\frac{\partial #1}{\partial #2}}
\newcommand{\deriv}[2]{\frac{d #1}{d #2}}

\title{Ising Spin Glass Under Continuous-Distribution Random
Magnetic Fields: Tricritical Points and Instability Lines}

\vskip \baselineskip

\author{Nuno Crokidakis}
\thanks{E-mail address: nuno@cbpf.br}

\author{Fernando D. Nobre}
\thanks{Corresponding author: E-mail address: fdnobre@cbpf.br}

\address{
Centro Brasileiro de Pesquisas F\'{\i}sicas \\
Rua Xavier Sigaud 150 \\
22290-180 \hspace{5mm} Rio de Janeiro - RJ \hspace{5mm} Brazil}

\date{\today}


\begin{abstract}
\noindent
The effects of random magnetic fields are considered in an Ising
spin-glass model defined in the limit of infinite-range interactions.
The probability distribution for the random magnetic fields is a double
Gaussian, which consists of two Gaussian 
distributions centered respectively, at $+H_{0}$ and $-H_{0}$, presenting
the same width $\sigma$. It is argued that such a distribution is more
appropriate for a theoretical description of real systems than its simpler
particular two 
well-known limits, namely 
the single Gaussian distribution ($\sigma \gg H_{0}$), and the bimodal one
($\sigma = 0$). 
The model is investigated by means of the replica
method, and phase diagrams are obtained within the replica-symmetric
solution. 
Critical frontiers exhibiting tricritical points occur for different values
of $\sigma$, with the possibility
of two tricritical points along the same critical frontier. To our
knowledge, it is the first  
time that such a behavior is verified for a spin-glass model in the
presence of a continuous-distribution random field, which represents a
typical situation of a real system.  
The stability of the replica-symmetric solution is analyzed, and the usual
Almeida-Thouless instability is verified for low temperatures. 
It is verified that, the higher-temperature tricritical point always
appears
in the region of stability of the replica-symmetric solution; a condition
involving the parameters $H_{0}$ and $\sigma$, 
for the occurrence of this tricritical point only, is obtained analytically.
Some of our
results are discussed in view of experimental measurements available in the
literature. 

\vskip \baselineskip


\end{abstract}
\maketitle


\noindent
\section{Introduction}

\vskip \baselineskip

Spin-glass systems 
\cite{dotsenkobook,nishimoribook,youngbook,fischerhertz,binderyoung}
continue to challenge many researchers in the area of magnetism. 
>From the theoretical point of view, its simplest version defined in terms
of Ising spin variables, the so-called 
Ising spin glass (ISG), represents one of the most fascinating problems 
in the physics of disordered magnets.
The ISG mean-field solution, based on the infinite-range-interaction model,
as proposed by Sherrington-Kirkpatrick (SK) \cite{sk}, presents a quite
nontrivial behavior. The correct low-temperature solution of the SK model 
is defined in terms of a continuous order-parameter function \cite{parisirsb}
(i.e., an infinite number of order parameters) associated with many
low-energy states, a procedure which is usually
denominated as replica-symmetry breaking (RSB). Furthermore, a transition
in the presence of an external magnetic field, known as the Almeida-Thouless
(AT) line \cite{at}, is found in the solution of the SK model: such a line
separates a low-temperature region, characterized by RSB, from a
high-temperature one, where a simple one-parameter solution, denominated as
replica-symmetric (RS) solution, is stable. 
The validity of the results of
the SK model for the description of more realistic systems, characterized
by short-range-interactions, 
represents a very polemic question \cite{youngbook}. Recent numerical
simulations claim the absence of an AT line in the three-dimensional
short-range ISG \cite{young04}, as well as along the non-mean-field region
of a one-dimensional ISG characterized by long-range 
interactions \cite{young05}. However, these results, obtained with rather
small lattice-size simulations, do not rule out the possibility of a
crossover to a different scenario at much larger lattice sizes, or also for
smaller fields (and/or temperatures). 
One candidate for alternative theory to the SK model is
the droplet model \cite{fisherhuse}, based on domain-wall
renormalization-group arguments for spin glasses
\cite{mcmillan84,braymoore87}. According to the droplet model, the
low-temperature phase of any \underline{finite-dimensional} short-range spin
glass should be described in terms of a single thermodynamic state
(together, of course, with its corresponding time-reversed counterpart), i.e.,
essentially a RS-type of solution. Many important features of the ISG still
deserve an appropriate understanding within the droplet-model scenario, and in
particular, the validity of this model becomes
questionable for increasing dimensionalities, where one expects the existence
of a finite upper critical dimension, above which the mean-field picture
should prevail.   

Some diluted antiferromagnets, like
${\rm Fe_{x}Zn_{1-x}F_{2}}$, ${\rm Fe_{x}Mg_{1-x}Cl_{2}}$ and 
${\rm Mn_{x}Zn_{1-x}F_{2}}$, have been the object of extensive experimental
research, due to their intriguing properties \cite{belangerreview}. 
These systems are able to
exhibit, within certain concentration ranges, random-field, spin-glass or
both behaviors, and in particular,  
the compounds ${\rm Fe_{x}Zn_{1-x}F_{2}}$ and 
${\rm Fe_{x}Mg_{1-x}Cl_{2}}$ are characterized by large crystal-field
anisotropies, in such a way that they may be reasonably well-described in
terms of Ising variables. Therefore, they are usually considered as good
physical realizations of the random-field Ising model (RFIM), or also of an
ISG. For the 
${\rm Fe_{x}Zn_{1-x}F_{2}}$, one gets a RFIM-like behavior for  
${\rm x} > 0.42$, an ISG
for ${\rm x} \sim 0.25$, whereas for intermediate concentrations
($0.25 < {\rm x} < 0.42$) one may observe both behaviors depending on
the magnitude of the applied external magnetic field [RFIM (ISG) for small
(large) magnetic fields], with a crossover between them
\cite{montenegro91,belanger91,rosalesrivera}.
In what concerns ${\rm Fe_{x}Mg_{1-x}Cl_{2}}$, one gets an ISG-like
behavior for ${\rm x} < 0.55$, whereas for $0.7 < {\rm x} < 1.0$ one has a
typical RFIM with a first-order transition turning into a continuous one
due to a change in the random fields
\cite{belangerreview,kushauerkleemann,kushauer}. 
Even though a lot of experimental data is available for these systems, they
still deserve an appropriate understanding, with only a few
theoretical models proposed for that purpose
\cite{barber,barbosa01,barbosa03,barbosa05,soares,nogueira,medeiros,vieira}.
Within the numerical-simulation technique, one has tried to take into
account the basic microscopic ingredients of such systems
\cite{barber,barbosa01,barbosa03,barbosa05}, whereas at the 
mean-field level, a
joint study of both ISG and RFIM models has been shown to be a very
promising approach \cite{soares,nogueira,medeiros,vieira}.   
 
In the present work we investigate the effects of random magnetic fields,
following a continuous probability distribution, 
in an ISG model. The model is 
considered in the limit of infinite-range interactions, and the 
probability distribution for the random magnetic fields is a double
Gaussian, which consists of a sum of two independent Gaussian
distributions. Such a 
distribution interpolates between the bimodal and the simple Gaussian
distributions, which are known to present distinct low-temperature
critical behavior, within the mean-field limit
\cite{soares,nogueira,medeiros,vieira}. It is argued that this distribution
is more appropriate for a theoretical description of diluted
antiferromagnets than the  
bimodal and Gaussian distributions. In particular, for given ranges of
parameters,  
this distribution presents two peaks, and satisfies the requirement of
effective random fields varying in both sign and magnitude, which comes out
naturally in the identification of the RFIM with diluted antiferromagnets
in the presence of a uniform field  
\cite{fishmanaharony,cardy}; this condition is not fulfilled by simple
discrete probability distributions, e.g., the bimodal one, which is
certainly very convenient from the theoretical point of view.   
Recently, the use a double-Gaussian distribution in the RFIM
\cite{nunocondmat} yielded interesting results, leading to a candidate
model to describe  
the change of a first-order transition into a continuous one that occurs in
${\rm Fe_{x}Mg_{1-x}Cl_{2}}$ \cite{belangerreview,kushauerkleemann,kushauer}.
The use of this distribution in the study of the present model should be 
relevant for ${\rm Fe_{x}Mg_{1-x}Cl_{2}}$ with concentrations $x<0.55$,
where the ISG behavior shows up. 
In the next section we study the SK
model in the presence of the above-mentioned random magnetic fields; a rich
critical 
behavior is presented, and in particular, one finds a critical frontier
that may present one, or even two, tricritical points. The instabilities of
the RS solution are also investigated, and AT lines presenting an
inflection point, in concordance with those measured in some diluted
antiferromagnets, are obtained. Finally, in section 4 we present our
conclusions.  

\vskip \baselineskip
\noindent
\section{The Ising Spin Glass in the Presence of a Random-Field}

\vskip \baselineskip

The infinite-range-interaction Ising spin-glass model, in the presence of
an external random magnetic field, may be defined in terms of the Hamiltonian

\begin{equation} \label{1}
\mathcal{H}=- \sum_{(i,j)}J_{ij}S_{i}S_{j} - \sum_{i}H_{i}S_{i}~, 
\end{equation}

\vskip \baselineskip
\noindent
where the sum $\sum_{(i,j)}$ applies to
all distinct pairs of spins $S_{i}=\pm 1$ ($i=1,2,...,N$). The interactions
$\{J_{ij}\}$ and the fields $\{H_{i}\}$ follow 
independent probability distributions, 

\begin{equation} \label{2}
P(J_{ij})=\left(\frac{N}{2\pi J^{2}}\right)^{1/2}
\exp\left[-\frac{N}{2J^{2}}\left(J_{ij}-\frac{J_{0}}{N}\right)^{2}\right]~,
\end{equation}

\vskip \baselineskip

\begin{equation} \label{3}
P(H_{i})=\frac{1}{2}\left(\frac{1}{2\pi \sigma^{2}}\right)^{1/2}
\left\{\exp\left[-\frac{(H_{i}-H_{0})^{2}}{2\sigma^{2}}\right]
+\exp\left[-\frac{(H_{i}+H_{0})^{2}}{2\sigma^{2}}\right]\right\}~. 
\end{equation}

\begin{figure}[t]
\begin{center}
\includegraphics[width=0.40\columnwidth,angle=0]{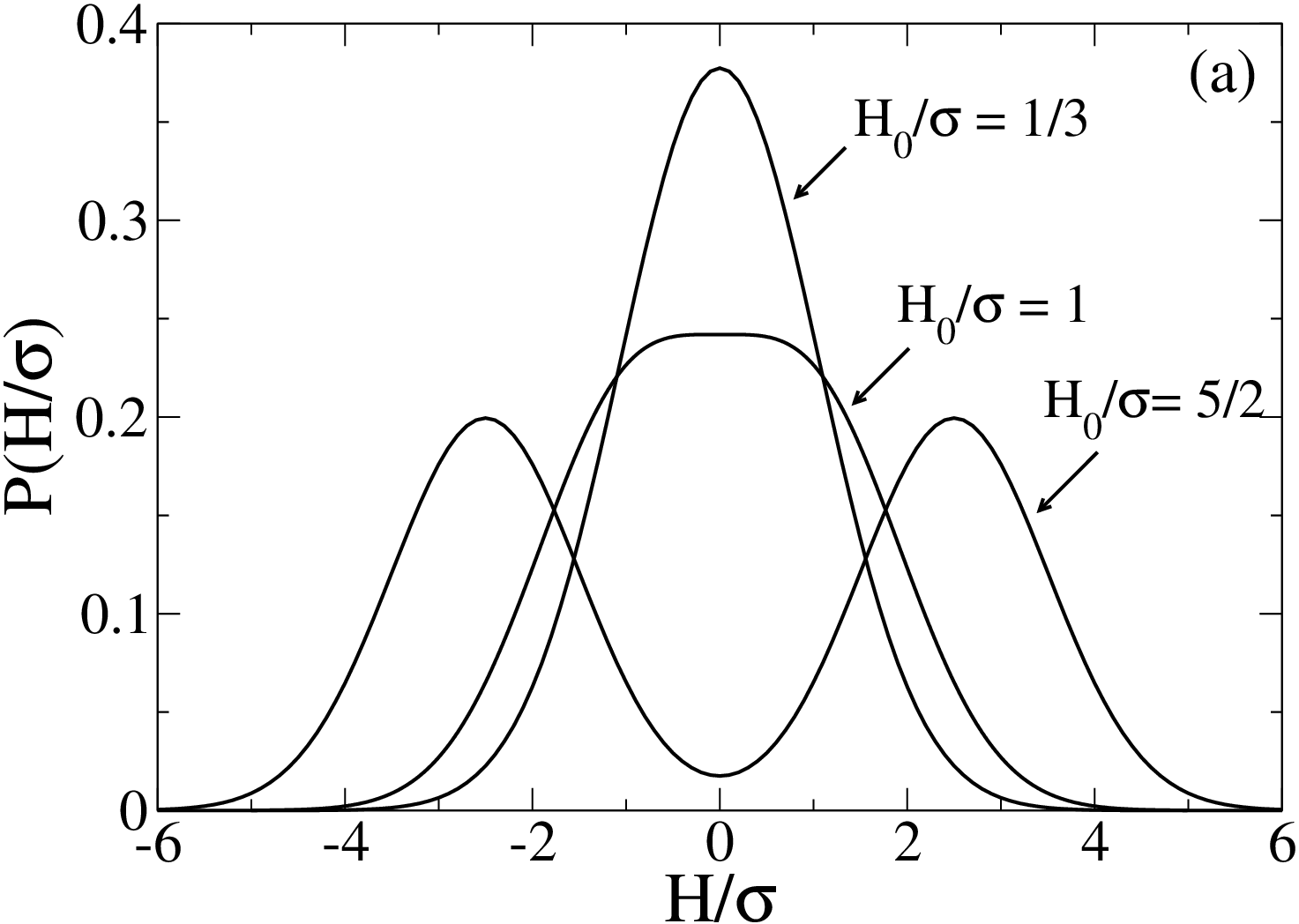}
\hspace{1.5cm}
\includegraphics[width=0.40\columnwidth,angle=0]{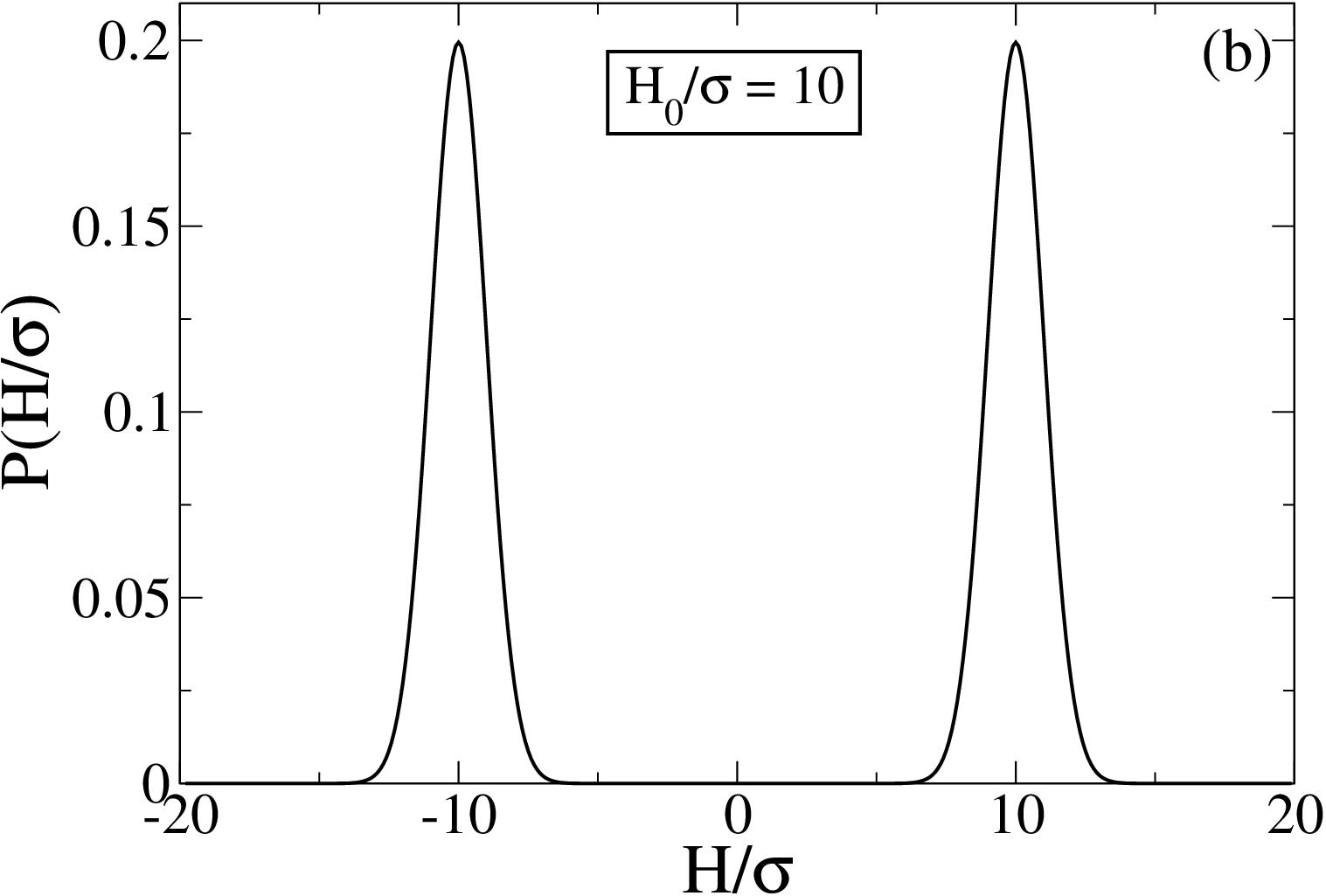}
\end{center}
\protect\caption{The probability distribution of Eq.~(\ref{3}) (the random
fields are scaled in units of $\sigma$) for
typical values of the ratio $H_{0}/\sigma$: (a) $(H_{0}/\sigma)=1/3,1,5/2$; 
(b) $(H_{0}/\sigma)=10$.}  
\label{fig1}
\end{figure}

\vskip \baselineskip
\noindent
The probability distribution for the fields $\{H_{i}\}$ is a 
double Gaussian and depends on two parameters, $H_{0}$ and
$\sigma$, modifying its form according to the ratio $H_{0}/\sigma$, as
exhibited in Fig.~1. Such a distribution is
double-peaked for $(H_{0}/\sigma)>1$, presents a single peak for
$(H_{0}/\sigma)<1$, changing its concavity at the 
origin when $(H_{0}/\sigma)=1$. Besides that, in the limit 
$\sigma \rightarrow 0$, one recovers a bimodal probability distribution.
It is important to notice that its kurtosis, 
$\kappa =  <H_{i}^{4}>_{H}/[3(<H_{i}^{2}>_{H})^{2}]$ 
[where $<\quad>_{H}$ denotes an average over $P(H_{i})$], varies from
$\kappa=1/3$ (bimodal limit) to $\kappa=1$ (Gaussian limit), approaching
unit only in the limit $(H_{0}/\sigma) \rightarrow 0$, in which case one gets
a perfect Gaussian distribution. For finite values of $H_{0}/\sigma$ one gets
$1/3<\kappa<1$, and in particular, for the cases exhibited in Fig.~1 one has
that $\kappa \approx 0.99 \ [(H_{0}/\sigma)=1/3]$, 
$\kappa \approx 0.83 \ [(H_{0}/\sigma)=1]$, 
$\kappa \approx 0.50 \ [(H_{0}/\sigma)=5/2]$, 
and $\kappa \approx 0.35 \ [(H_{0}/\sigma)=10]$.  

The free energy of a given disorder realization,
$F(\{J_{ij},H_{i}\})$, depends on two random variables, in such a way that
the average over the disorder $[\quad]_{J,H}$ may be written in terms of
independent integrals,  

\begin{equation} \label{4}
[F(\{J_{ij},H_{i}\})]_{J,H}=\int\prod_{(i,j)}[dJ_{ij}P(J_{ij})]
\prod_{i}[dH_{i}P(H_{i})]F(\{J_{ij},H_{i}\})~. 
\end{equation}

\vskip \baselineskip
\noindent
Now, one can make use of the replica method 
\cite{dotsenkobook,nishimoribook,fischerhertz,binderyoung} in order to
obtain the free energy per spin,

\begin{equation} \label{5}
-\beta f=\lim_{N \to \infty}\frac{1}{N}[\ln Z(\{J_{ij},H_{i}\})]_{J,H} 
= \lim_{N \to \infty}\lim_{n \to 0}\frac{1}{Nn}([Z^{n}]_{J,H}-1)~, 
\end{equation}

\vskip \baselineskip
\noindent
where $Z^{n}$ represents the partition function of the replicated system 
and $\beta = 1/(kT)$. Standart calculations lead to

\begin{equation} \label{6}
\beta f= -\frac{(\beta J)^{2}}{4}-\frac{(\beta\sigma)^{2}}{2}+ 
\lim_{n \to 0} \frac{1}{n}{\rm min} \; g(m^{\alpha},q^{\alpha\beta})~, 
\end{equation}

\vskip \baselineskip
\noindent
where

\begin{equation} \label{7}
g(m^{\alpha},q^{\alpha\beta})=\frac{\beta J_{0}}{2}\sum_{\alpha}
(m^{\alpha})^{2} + \frac{(\beta J)^{2}}{2}\sum_{(\alpha \beta)}
(q^{\alpha\beta})^{2}-\frac{1}{2}\ln{\rm Tr}_{\alpha}
\exp(\mathcal{H}_{{\rm eff}}^{+})-\frac{1}{2}
\ln{\rm Tr}_{\alpha}\exp(\mathcal{H}_{{\rm eff}}^{-})~, 
\end{equation}

\begin{equation} \label{8}
\mathcal{H}_{{\rm eff}}^{\pm}=\beta J_{0}\sum_{\alpha}m^{\alpha}S^{\alpha} 
+ (\beta\sigma)^{2}\sum_{(\alpha \beta)}S^{\alpha}S^{\beta} 
+ (\beta J)^{2}\sum_{(\alpha \beta)}q^{\alpha\beta}S^{\alpha}S^{\beta} 
\pm \beta H_{0}\sum_{\alpha}S^{\alpha}~. 
\end{equation}

\vskip \baselineskip
\noindent
In the equations above, the index $\alpha$ ($\alpha=1,2,...,n$) is a
replica label, ${\rm Tr}_{\alpha}$ represents a trace over the spin
variables of each replica, and 
$\sum_{(\alpha \beta)}$ denote sums over distinct pairs of replicas. 

The extrema of the functional $g(m^{\alpha},q^{\alpha \beta})$ give us the
equilibrium equations 

\begin{eqnarray}\label{9}
m^{\alpha} & = & \frac{1}{2}<S^{\alpha}>_{+}+\frac{1}{2}<S^{\alpha}>_{-}~, 
\\ \label{10}
q^{\alpha\beta} & = & \frac{1}{2}<S^{\alpha\beta}>_{+}
+\frac{1}{2}<S^{\alpha\beta}>_{-} \quad (\alpha \neq \beta)~,
\end{eqnarray}

\vskip \baselineskip
\noindent
where $<\quad>_{\pm}$ indicate thermal averages with respect to 
the ``effective Hamiltonians'' $\mathcal{H}_{eff}^{\pm}$ in 
Eq.~(\ref{8}).

Assuming the RS ansatz 
\cite{dotsenkobook,nishimoribook,fischerhertz,binderyoung}, i.e., 
$m^{\alpha}=m$ ($\forall \alpha$) \ and \
$q^{\alpha\beta}=q$ [$\forall (\alpha\beta)$], 
Eqs.~(\ref{6})--(\ref{10}) yield

\begin{eqnarray} \nonumber
\beta f & = & -\frac{(\beta J)^{2}}{4}(1-q)^{2}+\frac{\beta J_{0}}{2}m^{2}- 
\frac{1}{2}\frac{1}{\sqrt{2\pi}}
\int_{\infty}^{\infty}dze^{-z^{2}/2}\ln(2\cosh\xi^{+}) 
\\ \nonumber \\ \label{11}
& & -\frac{1}{2}\frac{1}{\sqrt{2\pi}}
\int_{\infty}^{\infty}dze^{-z^{2}/2}\ln(2\cosh\xi^{-})~, 
\\ \nonumber \\ \label{12}
m & = & \frac{1}{2}\frac{1}{\sqrt{2\pi}}\int_{-\infty}^{+\infty}dze^{-z^{2}/2}
\tanh\xi^{+}+\frac{1}{2}\frac{1}{\sqrt{2\pi}}
\int_{-\infty}^{+\infty}dze^{-z^{2}/2}\tanh\xi^{-}~,
\\ \nonumber \\ \label{13}
q & = & \frac{1}{2}\frac{1}{\sqrt{2\pi}}\int_{-\infty}^{+\infty}dze^{-z^{2}/2}
\tanh^{2}\xi^{+} + \frac{1}{2}\frac{1}{\sqrt{2\pi}}
\int_{-\infty}^{+\infty}dze^{-z^{2}/2}\tanh^{2}\xi^{-}~,
\end{eqnarray}

\vskip \baselineskip
\noindent
where

\begin{equation} \label{14}
\xi^{\pm}=\beta\left\{J_{0}m+JGz \ \pm H_{0}\right\}~,   
\end{equation}

\begin{equation} \label{14a}
G = \left[ q+\left(\frac{\sigma}{J}\right)^{2} \right]^{1/2}~. 
\end{equation}

\vskip \baselineskip
Although the spin-glass order parameter [Eq.~(\ref{13})] is always
induced by the random field, it may still contribute to a nontrivial
behavior. The RS solution is known to lead to an instability at low
temperatures, usually associated to this parameter,
occurring below the AT \cite{at} line, 

\begin{equation} \label{15}
\left(\frac{kT}{J}\right)^{2}=\frac{1}{2}
\frac{1}{\sqrt{2\pi}}\int_{-\infty}^{+\infty}dz
e^{-z^{2}/2}{\rm sech}^{4}\xi^{+} + \frac{1}{2}
\frac{1}{\sqrt{2\pi}}\int_{-\infty}^{+\infty}dz
e^{-z^{2}/2}{\rm sech}^{4}\xi^{-}~. 
\end{equation}

\vskip \baselineskip
Let us now present the phase diagrams of this model. Since the 
random field induces the parameter $q$, there is no spontaneous spin-glass
order, like the one found in the SK model. However, there is a phase
transition related to the 
magnetization $m$, in such a way that two phases are
possible within the RS solution, namely, the Ferromagnetic 
($m \neq 0$, $q \neq 0$) and the
Independent ($m = 0$, $q \neq 0$) ones. 
The critical frontier separating these two phases is obtained by solving
the equilibrium 
conditions, Eqs.~(\ref{12}) and (\ref{13}), whereas in the case of
first-order 
phase transitions, the free energy per spin, 
Eq.~(\ref{11}), will be analyzed. Expanding the magnetization
[Eq. (\ref{12})] in power series, 

\begin{equation} \label{16}
m = A_{1}(q)m + A_{3}(q)m^{3} + A_{5}(q)m^{5} + O(m^{7})~,
\end{equation}

\vskip \baselineskip
\noindent
where

\begin{eqnarray} \label{17}
A_{1}(q) &=& \beta J_{0}\{1-\rho_{1}(q)\}~, \\ \label{18}
A_{3}(q) &=& -\frac{(\beta J_{0})^{3}}{3}\{1-4\rho_{1}(q)
+3\rho_{2}(q)\}~,  \\ \label{19}
A_{5}(q) &=& \frac{(\beta J_{0})^{5}}{15}\{2-17\rho_{1}(q)+30\rho_{2}(q)
-15\rho_{3}(q)\}~,
\end{eqnarray}

\vskip \baselineskip
\noindent
and

\begin{equation} \label{20}
\rho_{k}(q)=\frac{1}{\sqrt{2\pi}}\int_{-\infty}^{+\infty}dz
e^{-z^{2}/2}\tanh^{2k}\beta J\left\{ Gz+\frac{H_{0}}{J} \right\}~. 
\end{equation}

\vskip \baselineskip
\noindent
The coefficients in Eqs.~(\ref{17})--(\ref{19}) depend on $q$, which in
its turn depends on $m$ through Eq.~(\ref{13}). In order to eliminate
this dependence, we expand Eq.~(\ref{13}) in powers of $m$, 

\begin{equation} \label{21}
q=q_{0}+(\beta J_{0})^{2}\frac{\Gamma}{1-(\beta J)^{2}\Gamma} \ 
m^{2}+O(m^{4})~, 
\end{equation}

\vskip \baselineskip
\noindent
with

\begin{equation} \label{22}
\Gamma=1-4\rho_{1}(q_{0})+3\rho_{2}(q_{0})~, 
\end{equation}

\vskip \baselineskip
\noindent
where $q_{0}$ corresponds to the solution of Eq.~(\ref{13}) for $m=0$.
Substituting Eq.~(\ref{21}) in the expansion of Eq.~(\ref{16}), 
one obtains the $m$-independent coefficients in the power expansion 
of the magnetization; in terms of the lowest-order coefficients, one gets, 

\begin{equation} \label{23}
m = A_{1}^{'}m + A_{3}^{'}m^{3} + O(m^{5})~, 
\end{equation}

\vspace{-10mm}

\begin{eqnarray} \label{24}
A_{1}^{'} &=& A_{1}(q_{0})~, \\ \label{25}
A_{3}^{'} &=& -\frac{(\beta J_{0})^{3}}{3}\left[\frac{1+2(\beta J)^{2}
\Gamma}{1-(\beta J)^{2}\Gamma}\right]\Gamma~.
\end{eqnarray}

\vskip \baselineskip
\noindent
The associated critical frontier is determined through the standard 
procedure, taking into account the spin-glass
order parameter [Eq.~(\ref{13})], as well. 
For continuous transitions,
$A_{1}^{'}=1$, with $A_{3}^{'}<0$, in such a way that one has to solve
numerically the equation $A_{1}^{'}=1$, together with Eq.~(\ref{13}) 
considering $m=0$.
If $A_{3}^{'}>0$, one may have first-order phase transitions, 
characterized by a discontinuity in the magnetization; in this case, the
critical frontier is found through a Maxwell construction, i.e., by
equating the free energies of the two phases, which should be solved
numerically together with Eqs.~(\ref{12}) and (\ref{13}) for each side
of the critical line.  
When both types of phase transitions are present, the continuous and
first-order critical frontiers meet at a tricritical point that
defines the limit of validity of the series expansion. The location of such
a point is determined by solving numerically equations $A_{1}^{'}=1$,
$A_{3}^{'}=0$, and Eq.~(\ref{13}) with $m=0$ [provided that the coefficient
of the next-order term in the expansion of Eq.~(\ref{23}) is negative,
i.e., $A_{5}^{'}<0$]. 

Considering the above-mentioned phases, the AT instability of
Eq.~(\ref{15}) splits each of them in 
two phases, in such a way that the phase diagram of this model may
present four phases, that are usually classified as
\cite{soares,nogueira,medeiros}: (i) Paramagnetic (P) ($m=0$; stability of
the RS solution); (ii) Spin-Glass (SG) ($m=0$; instability of the RS
solution); (iii) Ferromagnetic (F) ($m \neq 0$; stability of the RS
solution); (iv) Mixed Ferromagnetic (${\rm F}^{\prime}$) 
($m \neq 0$; instability of the RS solution). 

Even though in most cases the AT line is computed numerically, for large
values of $J_{0}$ [i.e., $J_{0}>>J$ and $J_{0}>>H_{0}$] and low
temperatures, one gets the following analytic asymptotic behavior,  

\begin{equation} \label{25a}
\frac{kT}{J}\cong \frac{2}{3}\frac{1}{\sqrt{2\pi}}\frac{1}{G}
\left\{\exp\left[-\frac{(J_{0}+H_{0})^{2}}{2J^{2}G^{2}}\right] +
\exp\left[-\frac{(J_{0}-H_{0})^{2}}{2J^{2}G^{2}}\right]\right\}~. 
\end{equation}

\begin{figure}[t]
\begin{center}
\includegraphics[width=0.40\textwidth,angle=0]{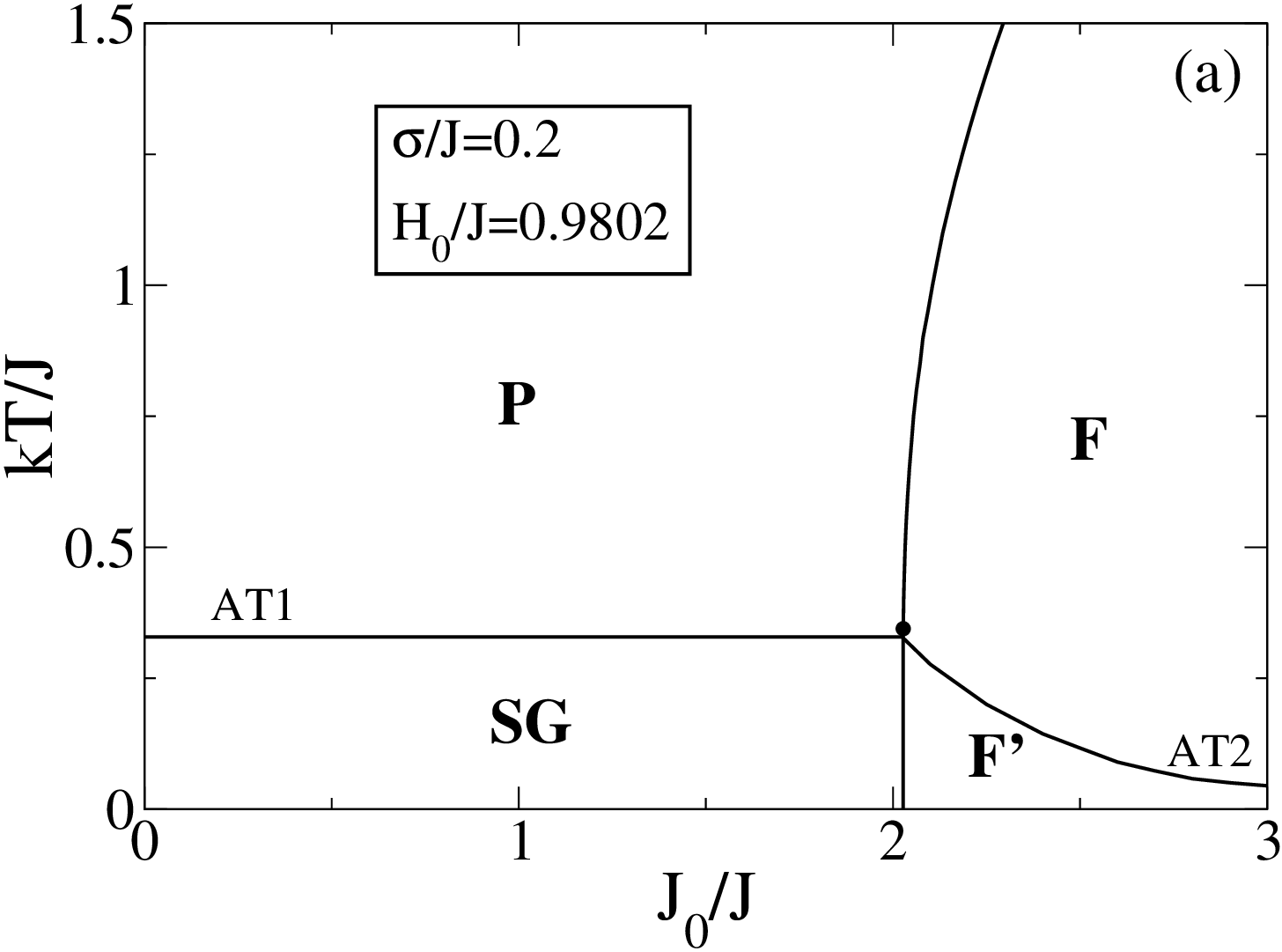}
\hspace{1.5cm}
\includegraphics[width=0.40\textwidth,angle=0]{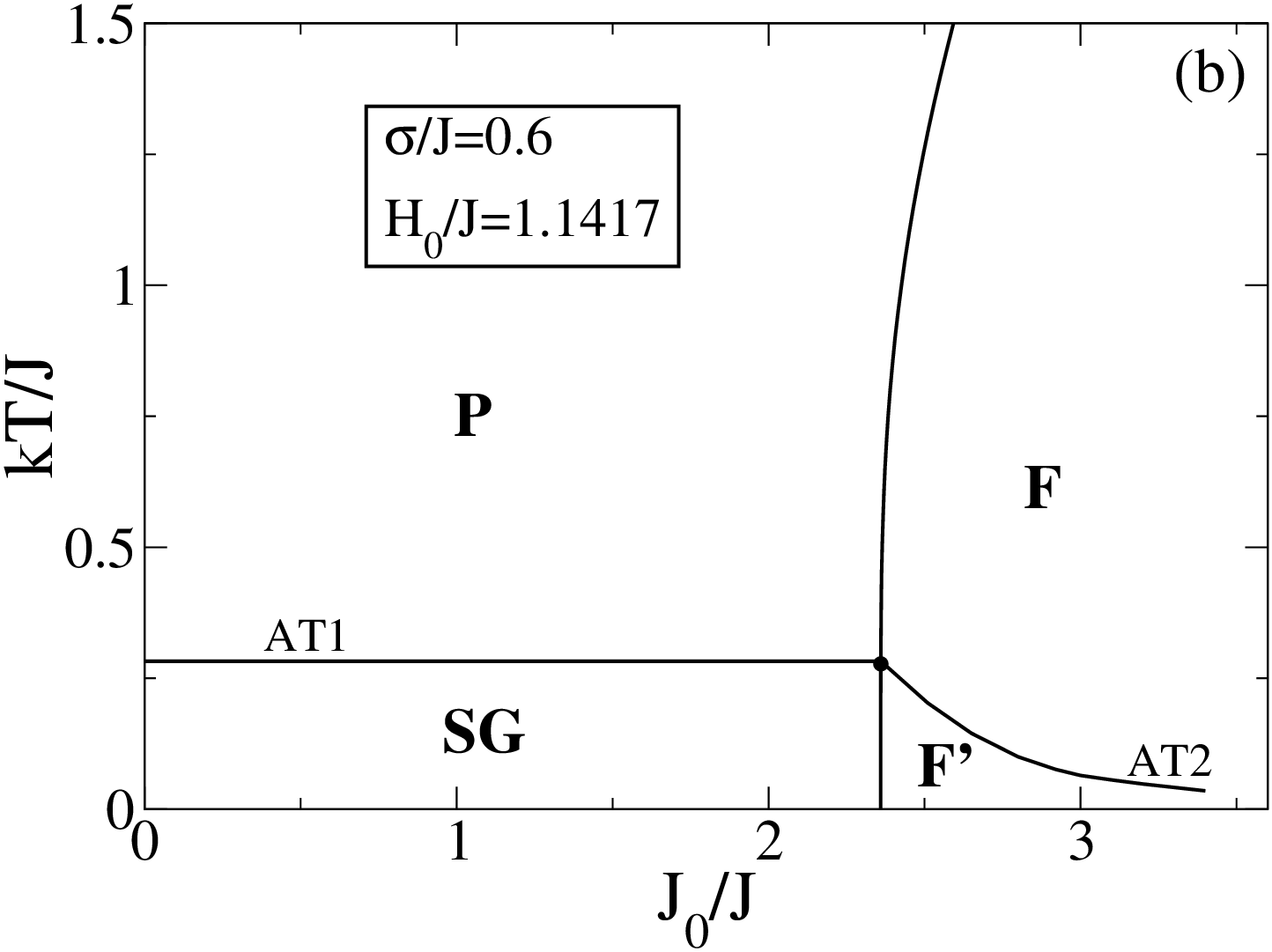}
\end{center}
\protect\caption{Phase diagrams of the infinite-range-interaction ISG in
the presence of a double-Gaussian random field; the phases are labelled
according to the definitions in the text. AT1 and AT2 denote AT 
lines, and all variables are scaled in 
units of $J$. Two typical examples [(a) $(\sigma/J)=0.2$; 
(b) $(\sigma/J)=0.6$]
are exhibited, for which there are single
points (represented by black dots) characterized by 
$A_{1}^{'}=1$ and $A_{3}^{'}=0$, defining the corresponding threshold
values $H_{0}^{(1)}(\sigma)$.}
\label{fig2}
\end{figure}

\vskip \baselineskip
For the particular case $\sigma=0$, i.e., the bimodal probability
distribution for the fields \cite{nogueira}, it was verified that the phase
diagrams of the model change qualitatively and quantitatively
for incresing values of $H_{0}$. Herein, we show that the phase diagrams of
the present model change according to the parameters of  
the distribution of random fields [Eq.~(\ref{3})], which may modify
drastically the critical line separating the regions with $m=0$ and 
$m \neq 0$, defined by the coefficients in Eq.~(\ref{23}). In particular,
one finds numerically a threshold value, $H_{0}^{(1)}(\sigma)$, for which
this line 
presents a single point characterized by $A_{1}^{'}=1$ and $A_{3}^{'}=0$;
all other points of this line represent continuous phase transitions,
characterized by $A_{1}^{'}=1$ and $A_{3}^{'}<0$. Typical examples of
this case are exhibited in Fig.~2, for the dimensionless ratios 
$(\sigma/J)=0.2$ and $(\sigma/J)=0.6$.
As will be seen in the next figures, for values of $H_{0}/J$ slightly
larger than  
$H_{0}^{(1)}(\sigma)/J$, this special point splits in two tricritical points,
whereas for values of $H_{0}/J$ smaller than 
$H_{0}^{(1)}(\sigma)/J$, this critical frontier is completely continuous.
Therefore, one may interpret the point for which 
$H_{0}=H_{0}^{(1)}(\sigma)$ as a collapse of two tricritical points.  
Such an unusual critical point is a characteristic of some
infinite-range-interaction spin-glasses in the presence of random
magnetic fields
\cite{nogueira,medeiros}, and to our knowledge, it has never been found 
in other magnetic models. 
>From Fig.~2, one notices that the threshold value 
$H_{0}^{(1)}(\sigma)/J$ increases for increasing values of $\sigma/J$,
although the corresponding ratio $H_{0}^{(1)}(\sigma)/\sigma$ decreases.
Apart from that, this peculiar critical point always occurs very close to
the onset of RSB; indeed, for  
$(\sigma/J)=0.6$, this point essentially coincides with the union of the two
AT lines (AT1 and AT2). At least for the range of ratios $\sigma/J$
investigated, this point never appeared 
below the AT lines, i.e., in the region of RSB. Therefore, an analysis that
takes into account RSB, will not modify the location of this point in these cases. 

\begin{figure}
\begin{center}
\includegraphics[width=0.40\textwidth,angle=0]{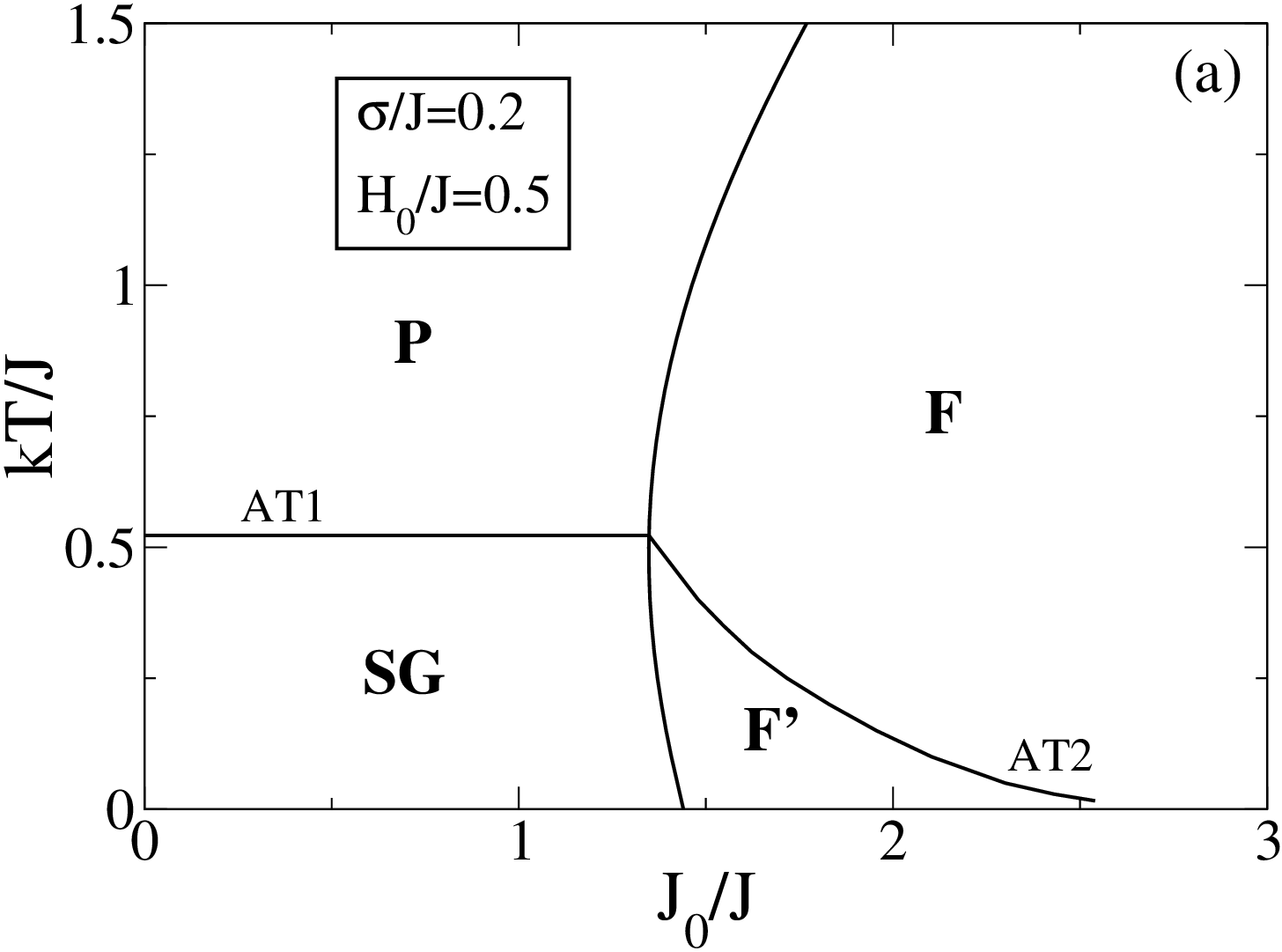}
\hspace{1.5cm}
\includegraphics[width=0.40\textwidth,angle=0]{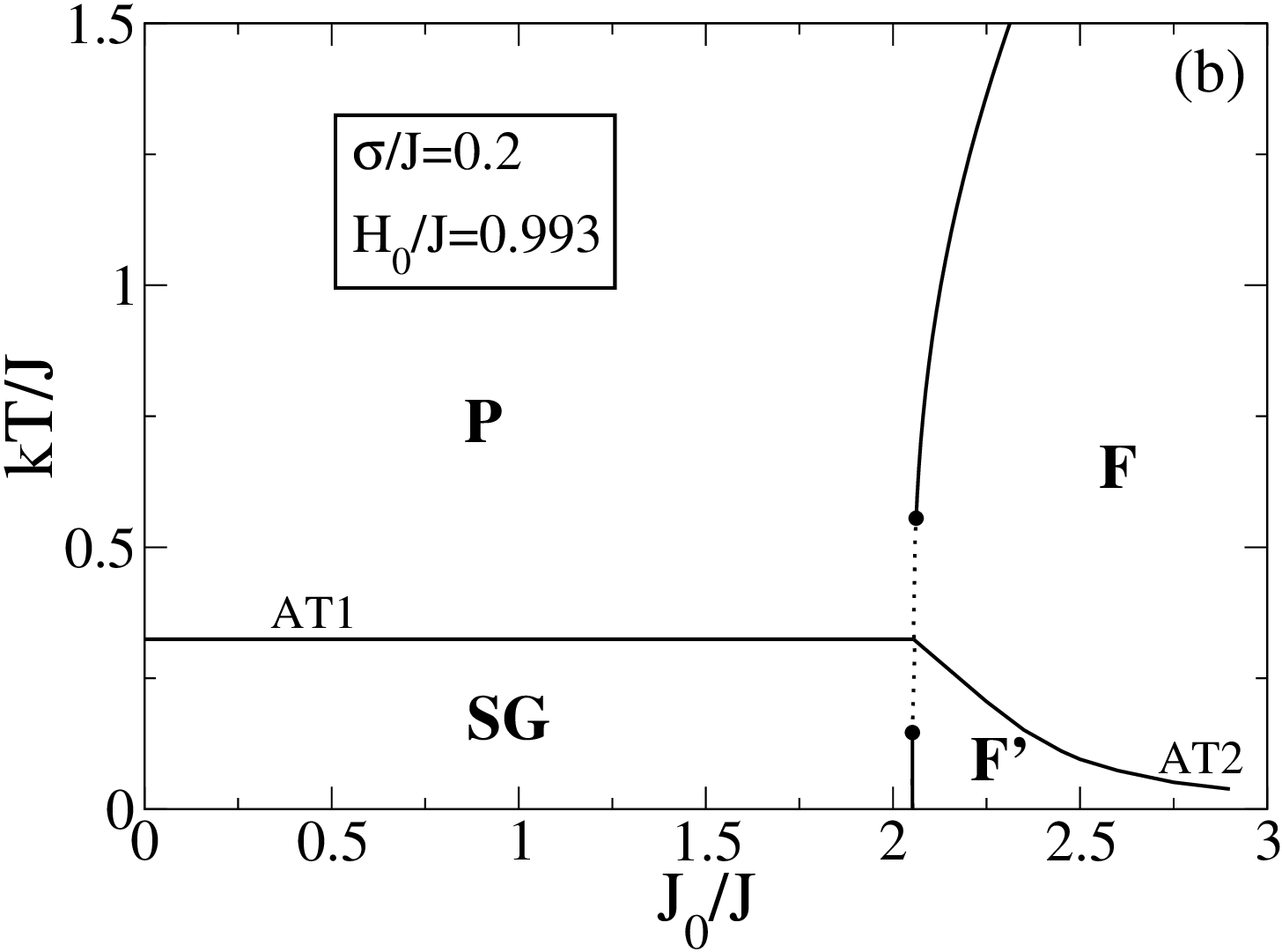}
\\
\vspace{1.0cm}
\includegraphics[width=0.40\textwidth,angle=0]{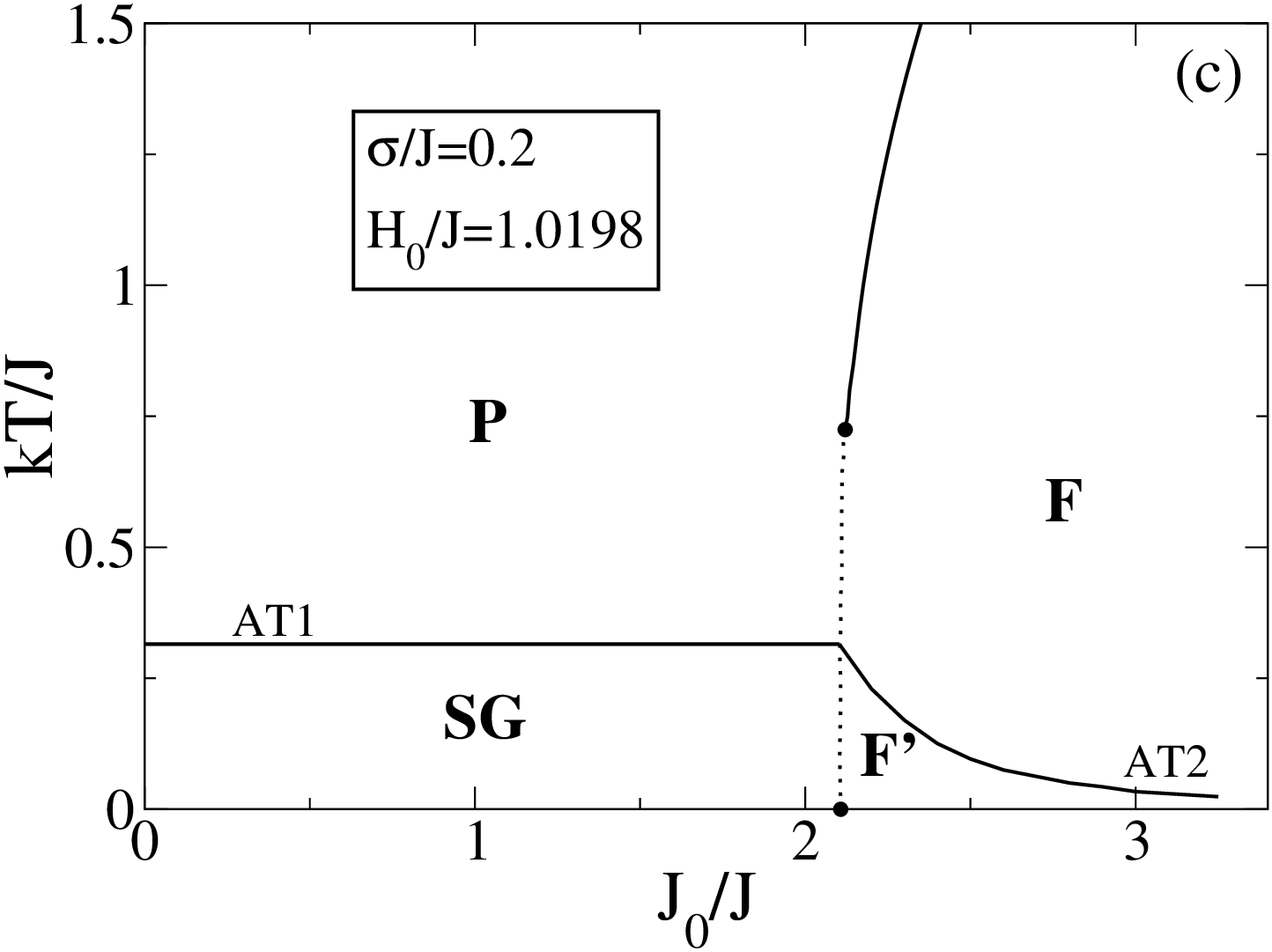}
\hspace{1.5cm}
\includegraphics[width=0.40\textwidth,angle=0]{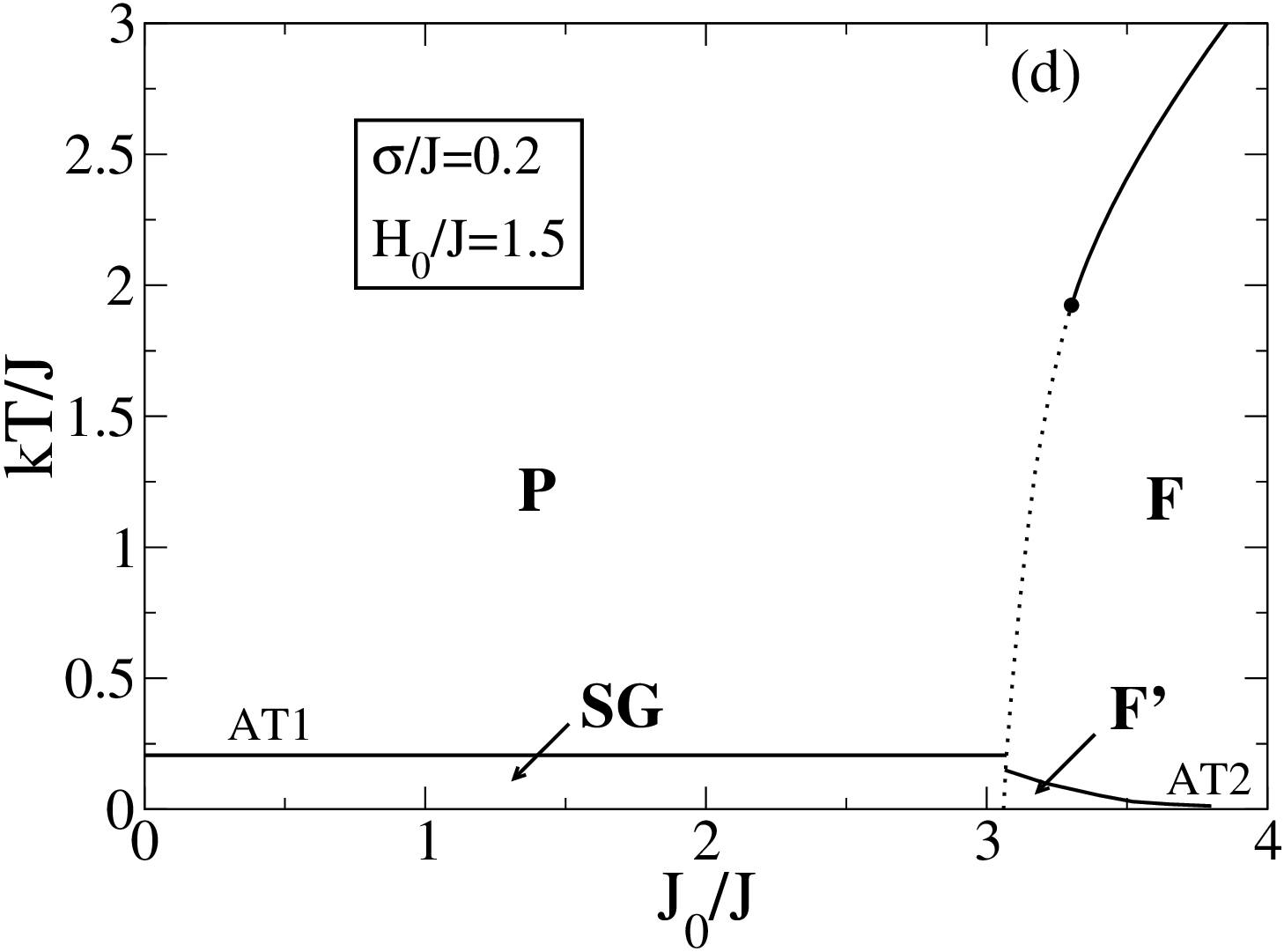}
\end{center}
\protect\caption{Phase diagrams of the infinite-range-interaction ISG in
the presence of a double-Gaussian random field with $(\sigma/J)=0.2$ and
typical values of $H_{0}/J$; the 
phases are labelled according to the definitions in the text. AT1 and AT2
denote AT lines, and all variables are scaled in 
units of $J$. 
By increasing the value of $H_{0}/J$, the phase diagram changes
both qualitatively and quantitatively and, particularly, the critical lines
separating the regions with $m=0$ and  
$m \neq 0$ are modified; along these critical frontiers,
the full (dotted) lines represent continuous (first-order) phase
transitions and the black dots denote tricritical points; for the values of 
$H_{0}/J$ chosen, one has: (a) continuous phase 
transitions; (b) two tricritical points at finite temperatures; 
(c) the lower tricritical point at 
zero temperature, defining the corresponding threshold value 
$H_{0}^{(2)}(\sigma)$; (d) a single tricritical point at finite
temperatures.}
\label{fig3}
\end{figure}

In Fig.~3 we exhibit phase diagrams for a fixed value of $\sigma$
($\sigma=0.2J$), and increasing values of $H_{0}$. 
In Fig.~3(a) we show the
case $(H_{0}/J)=0.5$, where one sees a phase diagram that looks like, at least
qualitatively, the one of the SK model; even though the random-field
distribution [cf. Eq.~(\ref{3})] is double-peaked (notice that
$(H_{0}/\sigma)=2.5$ in this case), 
the effects of such a field are not sufficient for a qualitative change in the
phase diagram of the model. As we have shown above [see Fig.~2(a)], 
qualitative changes only occur in the corresponding phase diagram for a ratio 
$(H_{0}^{(1)}(\sigma)/\sigma) \approx 5$, or higher. It is important to
remark that a  
tricritical point occurs in the corresponding RFIM for any 
$(H_{0}/\sigma) \ge 1$ 
\cite{nunocondmat}, in agreement with former general analyses
\cite{aharony78,andelman,galambirman}. If one associates the tricritical points 
that occur in the present model as reminiscents of the one in the RFIM, one
notices that such effects appear attenuated in the present model due to the
bond randomness, as predicted previously for short-range-interaction 
models \cite{aizenman,huiberker}.  
In Fig.~3(b) we present
the phase diagram for $(H_{0}/J)=0.993$; in this case, one observes two
finite-temperature tricritical points along the critical frontier that
separates the regions
with $m=0$ and $m \neq 0$. The higher-temperature point is
located in the region where the RS approximation is stable, and so, it will
not be affected by RSB effects; however, the lower-temperature tricritical
point, found in the region of instability of the RS solution, may change
under a RSB procedure. 
In Fig.~3(c) we exhibit another interesting
situation of the phase diagram of this model, for which the
lower-temperature tricritical point goes down to zero temperature, defining
a second threshold value, $H_{0}^{(2)}(\sigma)$. This threshold value was
calculated analytically, through a zero-temperature approach that follows
below, for arbitrary
values of $\sigma/J$. Above such a threshold,  
only the higher-temperature tricritical point (located in the region of
stability of the RS solution) exists; this is shown in
Fig.~3(d), where one considers a typical situation with
$H_{0}>H_{0}^{(2)}(\sigma)$. It is important to notice that in Fig.~3(d)
the two AT lines clearly do not meet at the critical frontier that
separates the regions 
with $m=0$ and $m \neq 0$; such an effect is a consequence of the phase
coexistence region, characteristic of first-order phase transitions, and
has already been observed in the SK model with a bimodal
random-field distribution \cite{nogueira}. The line AT1 is valid up to the
right end limit of the phase coexistence region, whereas AT2 remains valid
up to the left end limit of such a region; as a consequence of this, the
lines AT1 and AT2 do not meet at the corresponding
Independent-Ferromagnetic critical frontier.  

\begin{figure}[t]
\begin{center}
\includegraphics[width=0.40\textwidth,angle=0]{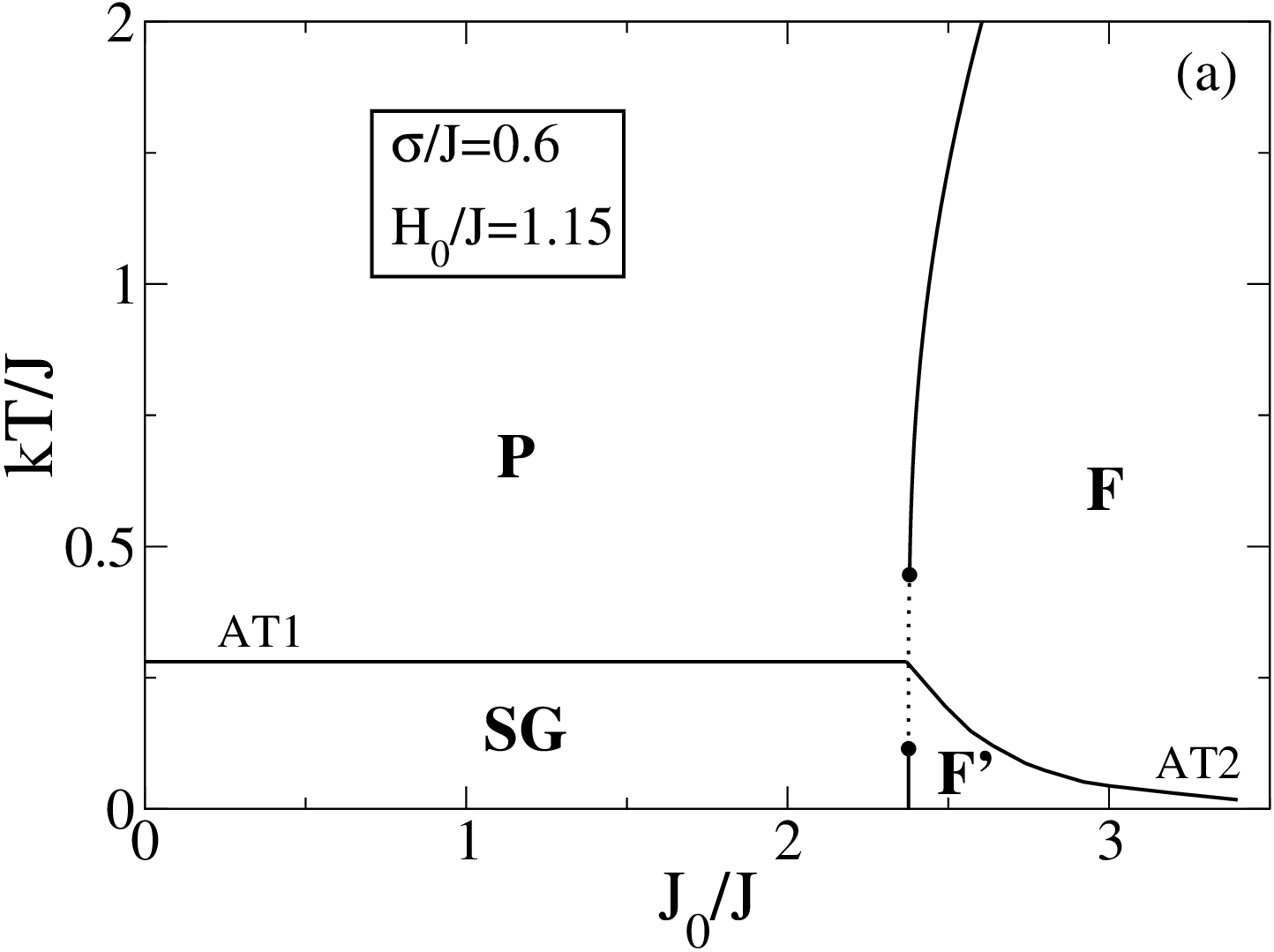}
\hspace{1.5cm}
\includegraphics[width=0.40\textwidth,angle=0]{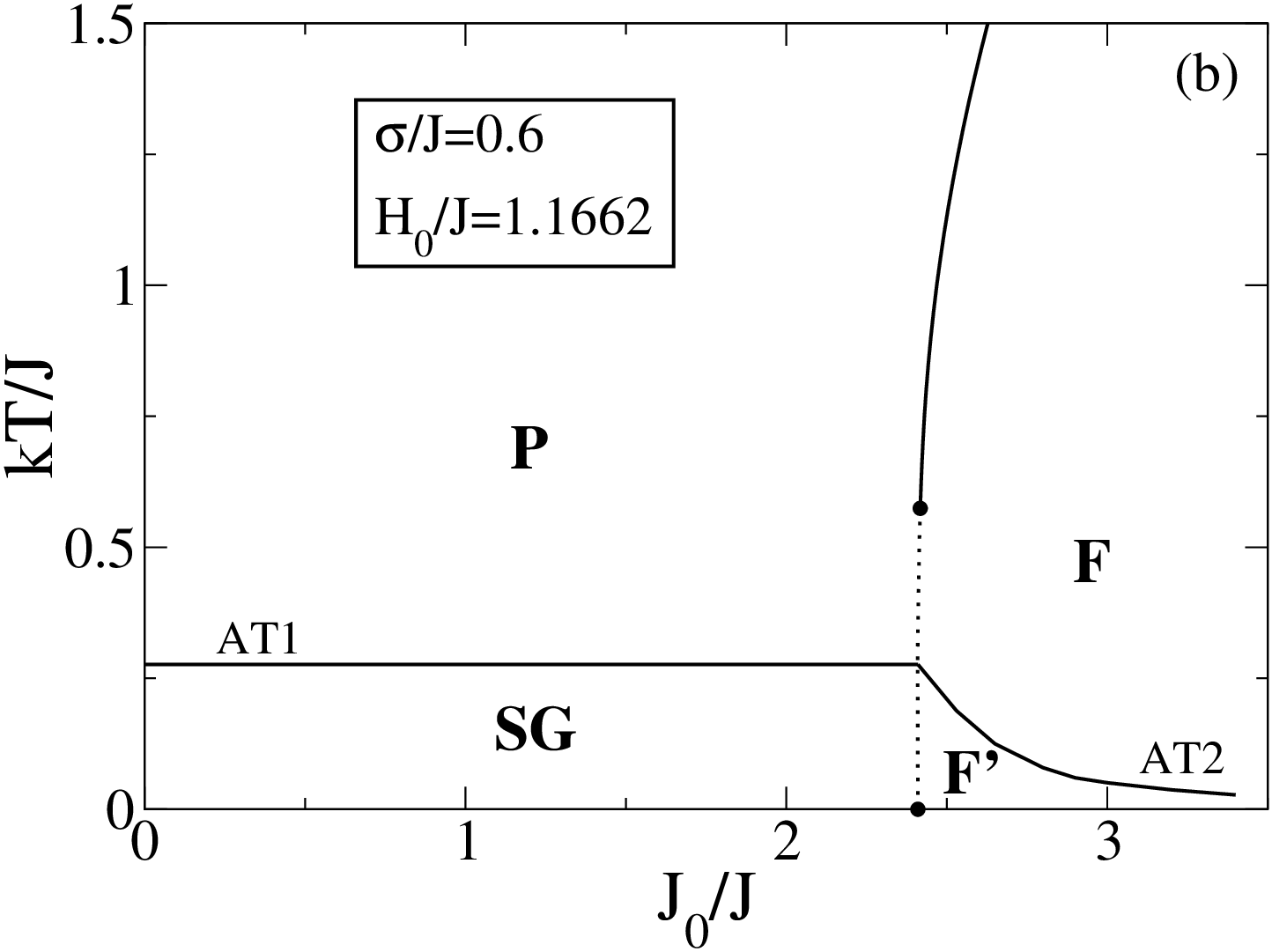}
\end{center}
\protect\caption{Phase diagrams of the infinite-range-interaction ISG in
the presence of a double-Gaussian random field with $(\sigma/J)=0.6$ and
typical values of $H_{0}/J$; the 
phases are labelled according to the definitions in the text. AT1 and AT2
denote AT lines, and all variables are scaled in 
units of $J$. 
Along the critical lines
separating the regions with $m=0$ and $m \neq 0$,
the full (dotted) lines represent continuous (first-order) phase
transitions and the black dots denote tricritical points; for the values of 
$H_{0}/J$ chosen, one has: (a) two tricritical points at finite
temperatures; (b) the lower tricritical point at
zero temperature, defining the corresponding threshold value 
$H_{0}^{(2)}(\sigma)$.}
\label{fig4}
\end{figure}

Additional phase diagrams are shown in Fig.~4, where we
exhibit two typical cases for the random-field width $(\sigma/J)=0.6$. 
In Fig.~4(a) we show the
equivalent of Fig.~3(b), where two tricritical points appear at finite
temperatures; now one gets qualitatively 
a similar effect, but with a random-field distribution characterized by  
a smaller ratio $H_{0}/\sigma$. From the quantitative point of view, the
following changes occur, in the critical  
frontier Independent-Ferromagnetic, due to an increase in $\sigma/J$: 
(i) such a critical frontier moves to higher values of $J_{0}/J$,
leading to an enlargement of the Independent phase [corresponding to the
region occupied by the P and SG phases of Fig.~4(a)]; (ii) the two
tricritical 
points are shifted to lower temperatures. 
In Fig.~4(b) we present the situation of a
zero-temperature tricritical point, defining the corresponding threshold
value $H_{0}^{(2)}(\sigma)$; once again, one gets a physical situation
similar to the one exhibited in Fig.~3(c), but with a much smaller ratio 
$H_{0}/\sigma$. 
Qualitatively similar effects may be also observed for other values of 
$\sigma$, but with different threshold values, $H_{0}^{(1)}(\sigma)$ and 
$H_{0}^{(2)}(\sigma)$. We have noticed that such 
threshold values increase with $\sigma/J$, even though
one requires less-pronounced double-peaked distributions [i.e., smaller
values for the ratios $H_{0}/\sigma$] in such a way to get significant
changes in the standard SK model phase diagrams [as can be seen in Figs. 2, 
3(c), and 4(b)]. 

\begin{figure}[t]
\begin{center}
\includegraphics[width=0.49\textwidth,angle=0]{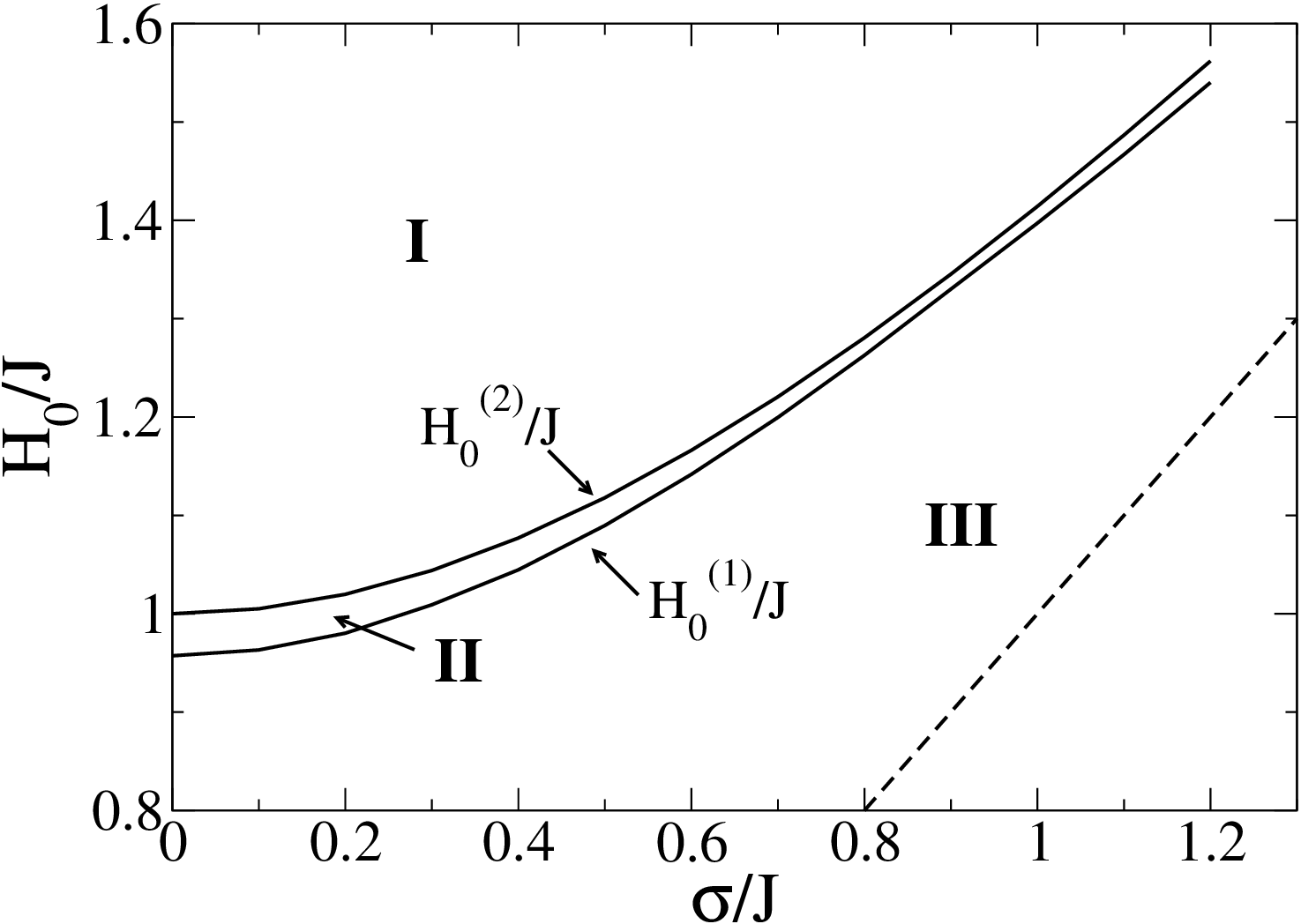}
\end{center}
\protect\caption{Evolution of the threshold values $H_{0}^{(1)}(\sigma)$
(lower curve) and $H_{0}^{(2)}(\sigma)$ (upper curve) with the
width $\sigma$ (all variables are scaled in units of $J$). 
Three distinct regions (I,II, and III) are
shown, concerning the existence of tricritical points and first-order phase
transitions along the Independent-Ferromagnetic critical frontier. The
dashed straight line 
corresponds to $H_{0}=\sigma$, above which one has a tricritical point in
the corresponding RFIM \cite{nunocondmat}.} 
\label{fig5}
\end{figure}

The evolution of 
the threshold values $H_{0}^{(1)}(\sigma)$ and $H_{0}^{(2)}(\sigma)$ 
with the dimensionless width  
$\sigma/J$ is exhibited in Fig.~5. One notices three distinct regions in
what concerns the existence of tricritical points and first-order phase
transitions along the Independent-Ferromagnetic critical frontier.
Throughout region I [defined
for $H_{0}>H_{0}^{(2)}(\sigma)$] a first-order phase transition occurs at
finite temperatures and reaches the zero-temperature axis; a single
tricritical point is found at finite temperatures [a typical example is
shown in  
Fig.~3(d)]. In region II [defined for
$H_{0}^{(1)}(\sigma)<H_{0}<H_{0}^{(2)}(\sigma)$] one finds two
finite-temperature tricritical points, with a first-order line between them
[typical examples are exhibited in Figs.~3(b) and 4(a)]. Along region III  
[$H_{0}<H_{0}^{(1)}(\sigma)$] one has a completely continuous 
Independent-Ferromagnetic critical frontier [like in Fig.~3(a)]. The dashed
straight line 
corresponds to   
$H_{0}=\sigma$, which represents the threshold for the existence of a
tricritical point in the corresponding RFIM \cite{nunocondmat}. 
Hence, if one associates the occurrence of tricritical points in the
present model with those of 
the RFIM, one notices that such effects are attenuated due to the bond
randomness, in agreement with Refs.~\cite{aizenman,huiberker}; herein, the
bond randomness introduces a spin-glass order parameter, 
in such a way that one needs
stronger values of $H_{0}/J$ for these tricritical points to occur. 

\begin{figure}[t]
\begin{center}
\includegraphics[width=0.49\textwidth,angle=0]{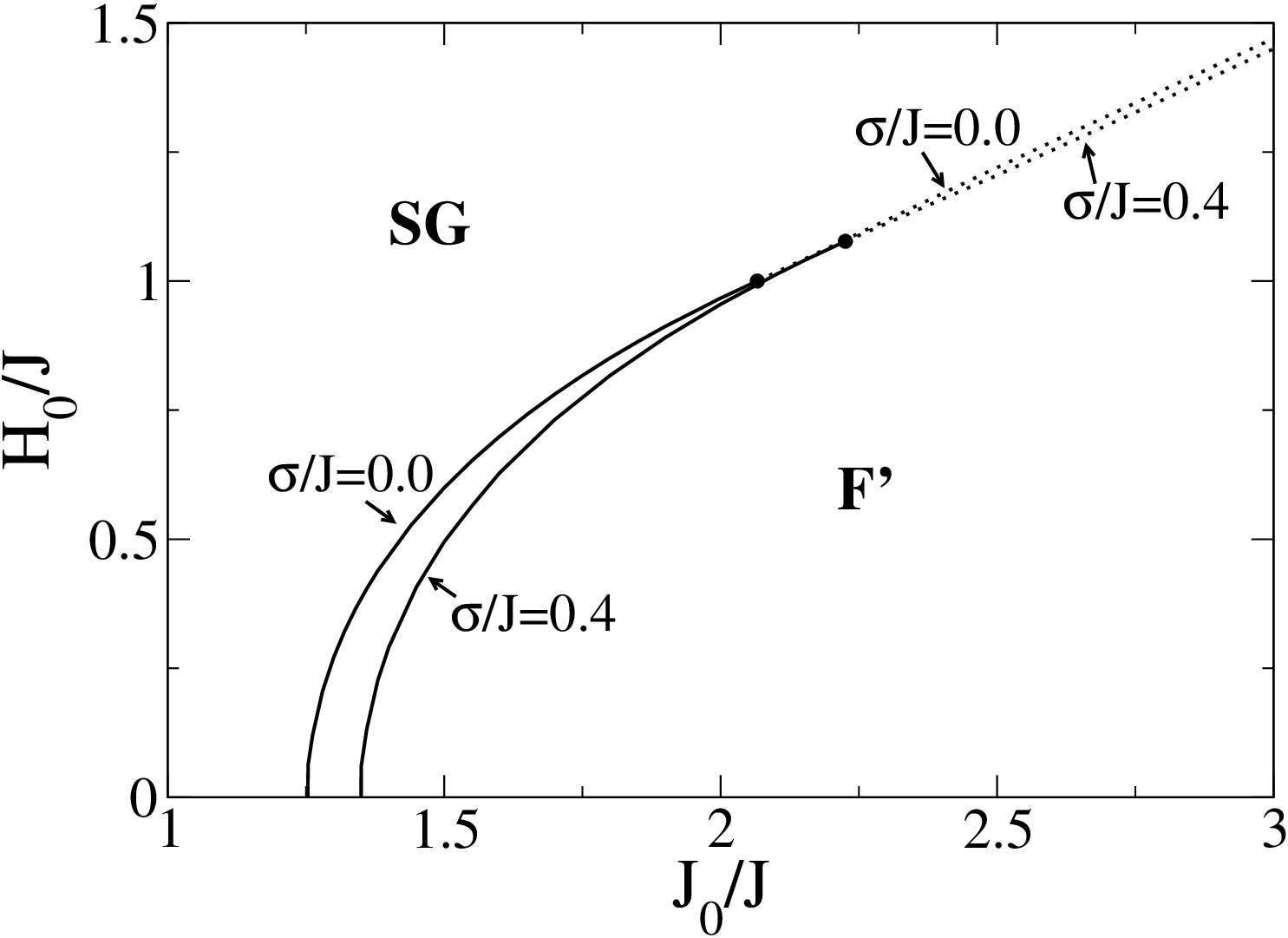}
\end{center}
\protect\caption{The zero-temperature phase diagram $H_{0}$ versus $J_{0}$ (in
units of J) for two typical values of the dimensionless width $\sigma/J$. The
critical frontiers separating the phases SG and ${\rm F}^{\prime}$ is 
continuous for small values of  
$H_{0}/J$ 
(full lines) and become first-order for higher values of $H_{0}/J$ (dotted
lines); the black dots denote tricritical points. Although the two critical
frontiers become very close near the tricritical points, they do not cross each
other; the tricritical point located at a higher value of $J_{0}/J$ corresponds
to the higher dimensionless width $\sigma/J$.} 
\label{fig6}
\end{figure}

Let us now consider the phase diagram of the model at zero temperature; in
this case, the spin-glass order parameter is trivial ($q=1$), in such a way
that the free energy and magnetization become,  

\begin{eqnarray} \nonumber
f & = & -\frac{J_{0}}{2}m^{2}- \frac{H_{0}}{2}\left[{\rm
erf}\left(\frac{J_{0}m+H_{0}}{JG_{0}\sqrt{2}}\right) - {\rm
erf}\left(\frac{J_{0}m-H_{0}}{JG_{0}\sqrt{2}}\right)\right] \\ \label{26}
& &\mbox{}-\frac{J}{\sqrt{2\pi}}G_{0}
\left\{\exp\left[-\frac{(J_{0}m+H_{0})^{2}}{2J^{2}G_{0}^{2}}\right] +
\exp\left[-\frac{(J_{0}m-H_{0})^{2}}{2J^{2}G_{0}^{2}}\right]\right\}~,
\end{eqnarray}

\begin{eqnarray}
m & = & \frac{1}{2}{\rm erf} \left(\frac{J_{0}m+H_{0}}{JG_{0}\sqrt{2}}\right) +
\frac{1}{2}{\rm erf}\left(\frac{J_{0}m-H_{0}}{JG_{0}\sqrt{2}}\right)~, 
\label{27}
\end{eqnarray}

\vskip \baselineskip
\noindent
where

\begin{equation} \label{28}
G_{0} = \left[ 1+\left(\frac{\sigma}{J}\right)^{2} \right]^{1/2}~. 
\end{equation}

\vskip \baselineskip
\noindent
Using a procedure similar to the one applied for finite temperatures, one
may expand Eq.~(\ref{27}) in powers of $m$, 

\begin{equation}
m=a_{1}m+a_{3}m^{3}+a_{5}m^{5}+O(m^{7})~, \label{29}
\end{equation}

\vskip \baselineskip
\noindent
where

\vskip \baselineskip
\noindent
\begin{eqnarray}\label{30}
a_{1} &=&
\sqrt{\frac{2}{\pi}}\frac{1}{G_{0}}\left(\frac{J_{0}}{J}\right)
\exp\left(-\frac{H_{0}^{2}}{2J^{2}G_{0}^{2}}\right)~,  \\ \label{31}
a_{3} &=&
\frac{1}{6}\sqrt{\frac{2}{\pi}}\frac{1}{G_{0}^{3}}
\left(\frac{J_{0}}{J}\right)^{3}
\left\{\frac{1}{G_{0}^{2}}\left(\frac{H_{0}}{J}\right)^{2}-1\right\}
\exp\left(-\frac{H_{0}^{2}}{2J^{2}G_{0}^{2}}\right)~,  \\ \label{32}
a_{5} &=&
\frac{1}{120}\sqrt{\frac{2}{\pi}}\frac{1}{G_{0}^{5}}
\left(\frac{J_{0}}{J}\right)^{5}
\left\{\frac{1}{G_{0}^{4}}\left(\frac{H_{0}}{J}\right)^{4} -
\frac{6}{G_{0}^{2}}\left(\frac{H_{0}}{J}\right)^{2}+3\right\}
\exp\left(-\frac{H_{0}^{2}}{2J^{2}G_{0}^{2}}\right)~.
\end{eqnarray}

\begin{figure}
\begin{center}
\includegraphics[width=0.40\textwidth,angle=0]{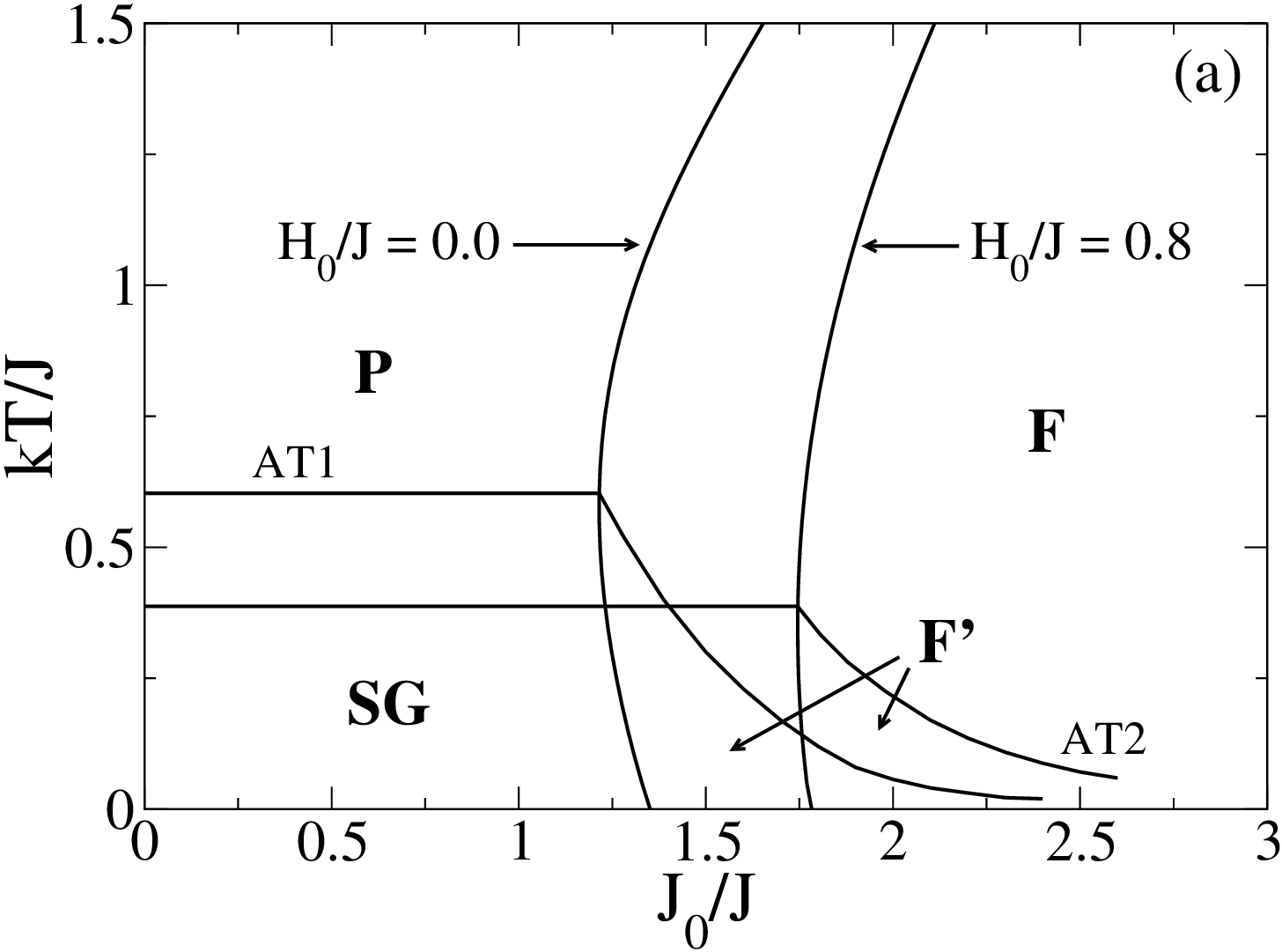}
\hspace{1.5cm}
\includegraphics[width=0.40\textwidth,angle=0]{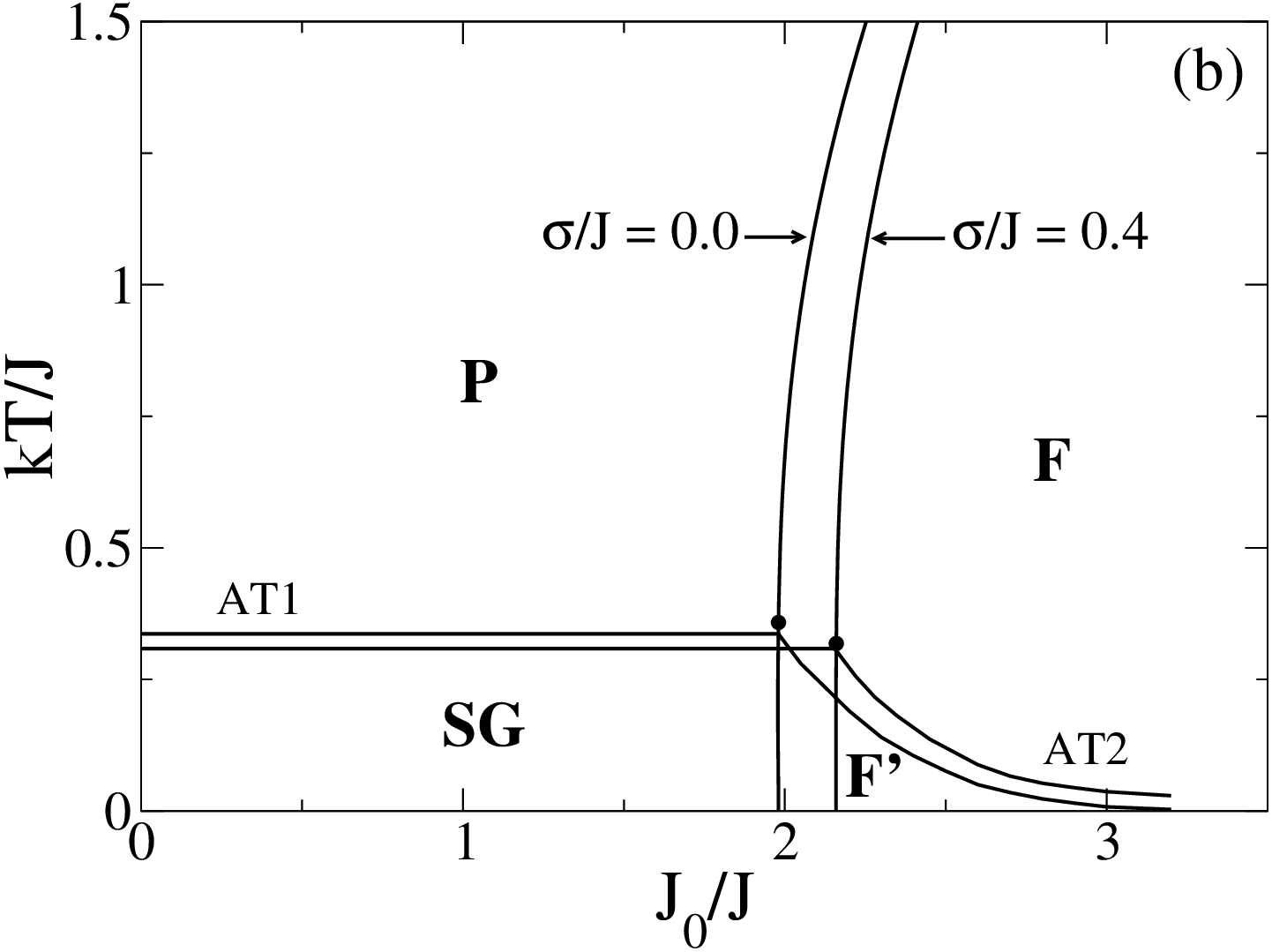}
\\
\vspace{1.0cm}
\includegraphics[width=0.40\textwidth,angle=0]{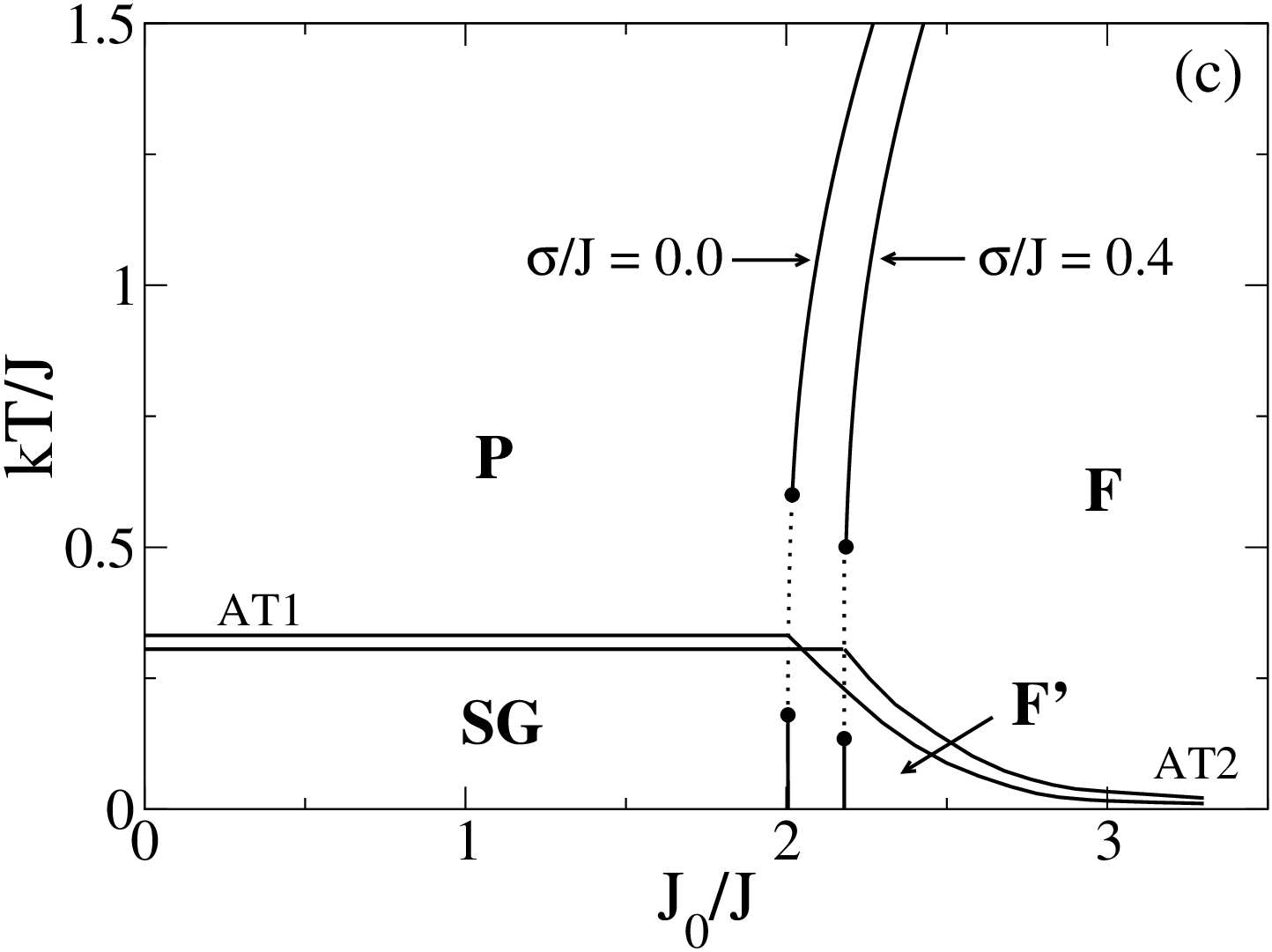}
\hspace{1.5cm}
\includegraphics[width=0.40\textwidth,angle=0]{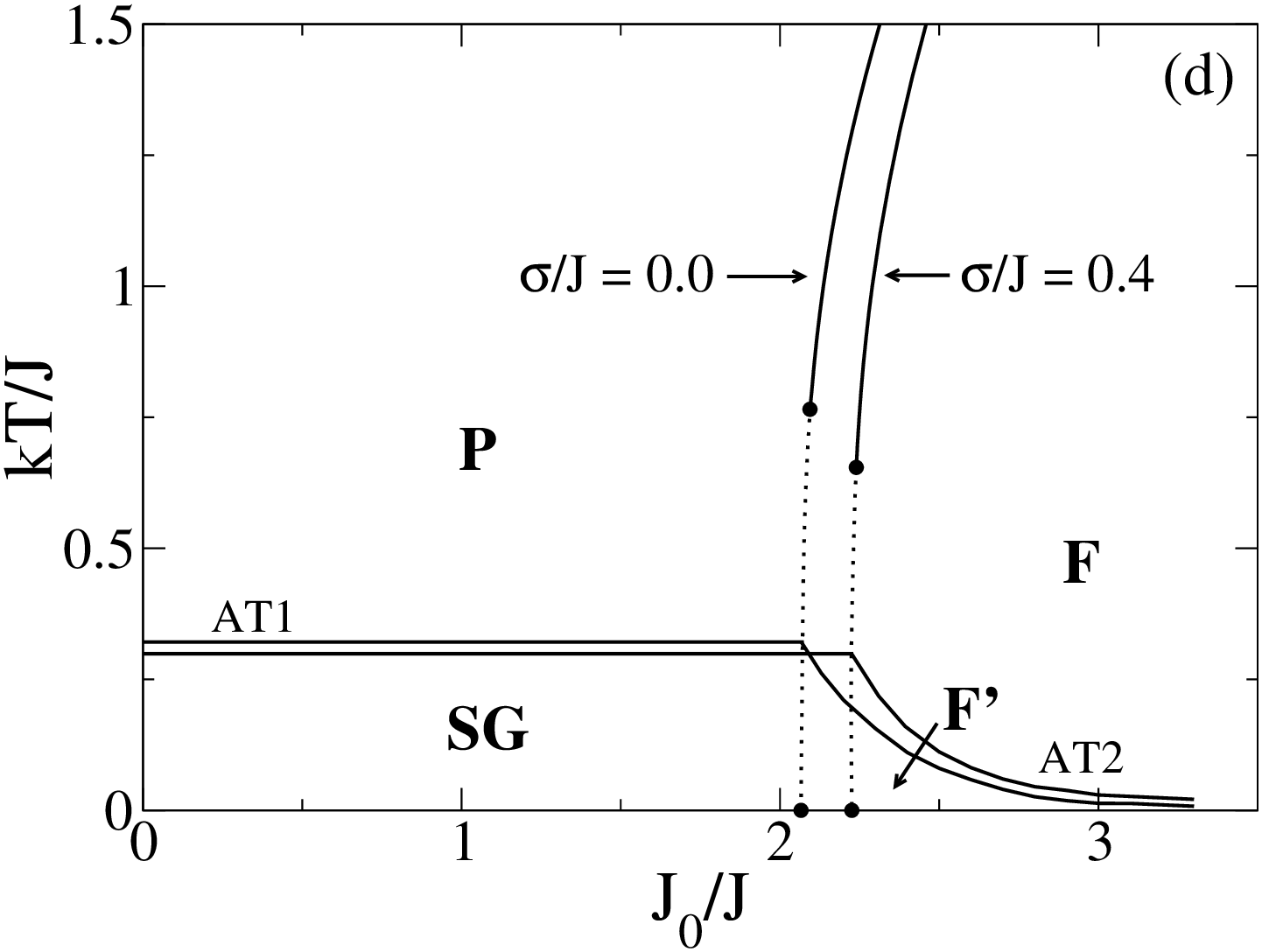}
\end{center}
\protect\caption{Typical phase diagrams of the infinite-range-interaction 
ISG in
the presence of a double-Gaussian random field with $(\sigma/J)=0.4$ are
compared with those already known for some particular cases. Comparisons of
qualitatively similar phase diagrams are presented, essentially in what
concerns the critical frontier that separates the regions with $m=0$ and
$m\neq0$. 
(a) Phase diagrams for the single Gaussian  
[$(H_{0}/J)=0.0$ and $(\sigma/J)=0.4$] and the double Gaussian
[$(H_{0}/J)=0.8$ and $(\sigma/J)=0.4$] distributions for the random fields.
(b) Phase diagrams for the bimodal  
[$(H_{0}/J)=0.9573$] and the double Gaussian
[$(H_{0}/J)=1.0447$] distributions for the random fields.
(c) Phase diagrams for the bimodal  
[$(H_{0}/J)=0.97$] and the double Gaussian
[$(H_{0}/J)=1.055$] distributions for the random fields.
(d) Phase diagrams for the bimodal  
[$(H_{0}/J)=1.0$] and the double Gaussian
[$(H_{0}/J)=1.077$] distributions for the random fields.
The phases are labelled according to the definitions in the text. 
AT1 and AT2
denote AT lines, and all variables are scaled in 
units of $J$.} 
\label{fig7}
\end{figure}

\vskip \baselineskip
\noindent
For $[{H_{0}}/(JG_{0})]^{2}<1$ [i.e., $a_{3}<0$], we have a continuous
critical frontier given by $a_{1}=1$,

\begin{equation}
\frac{J_{0}}{J}=\sqrt{\frac{\pi}{2}} \ G_{0} \exp\left[\frac{H_{0}^{2}}
{2J^{2}G_{0}^{2}}\right]~. 
\label{33}
\end{equation}

\vskip \baselineskip
\noindent
This continuous critical frontier ends at a tricritical point ($a_{3}=0$), 

\begin{equation}
\frac{1}{G_{0}^{2}}\left(\frac{H_{0}}{J}\right)^{2}=1 \quad \Rightarrow \quad
\frac{H_{0}}{J}  \equiv \frac{H_{0}^{(2)}}{J} 
=\left[1+\left(\frac{\sigma}{J}\right)^{2}\right]^{1/2}~, \label{34}
\end{equation}

\vskip \baselineskip
\noindent
which may be substituted in Eq.~(\ref{33}) to give

\begin{equation}
\frac{J_{0}}{J}=\sqrt{\frac{\pi e}{2}}\left[1+\left(\frac{\sigma}
{J}\right)^{2}\right]^{1/2}~. \label{35}
\end{equation}

\vskip \baselineskip
\noindent
Hence, Eqs.~(\ref{34}) and (\ref{35}) yield the coordinates of the
tricritical point at zero temperature. In addition to that, the result of
Eq.~(\ref{34}) corresponds to the exact threshold value
$H_{0}^{(2)}(\sigma)$ (as exhibited in Fig.~5).  
The above results are represented in the zero-temperature phase diagram
shown in Fig.~6, where one finds a single critical frontier separating the
phases SG and  
${\rm F}^{\prime}$. 

\begin{figure}[t]
\begin{center}
\includegraphics[width=0.40\columnwidth,angle=0]{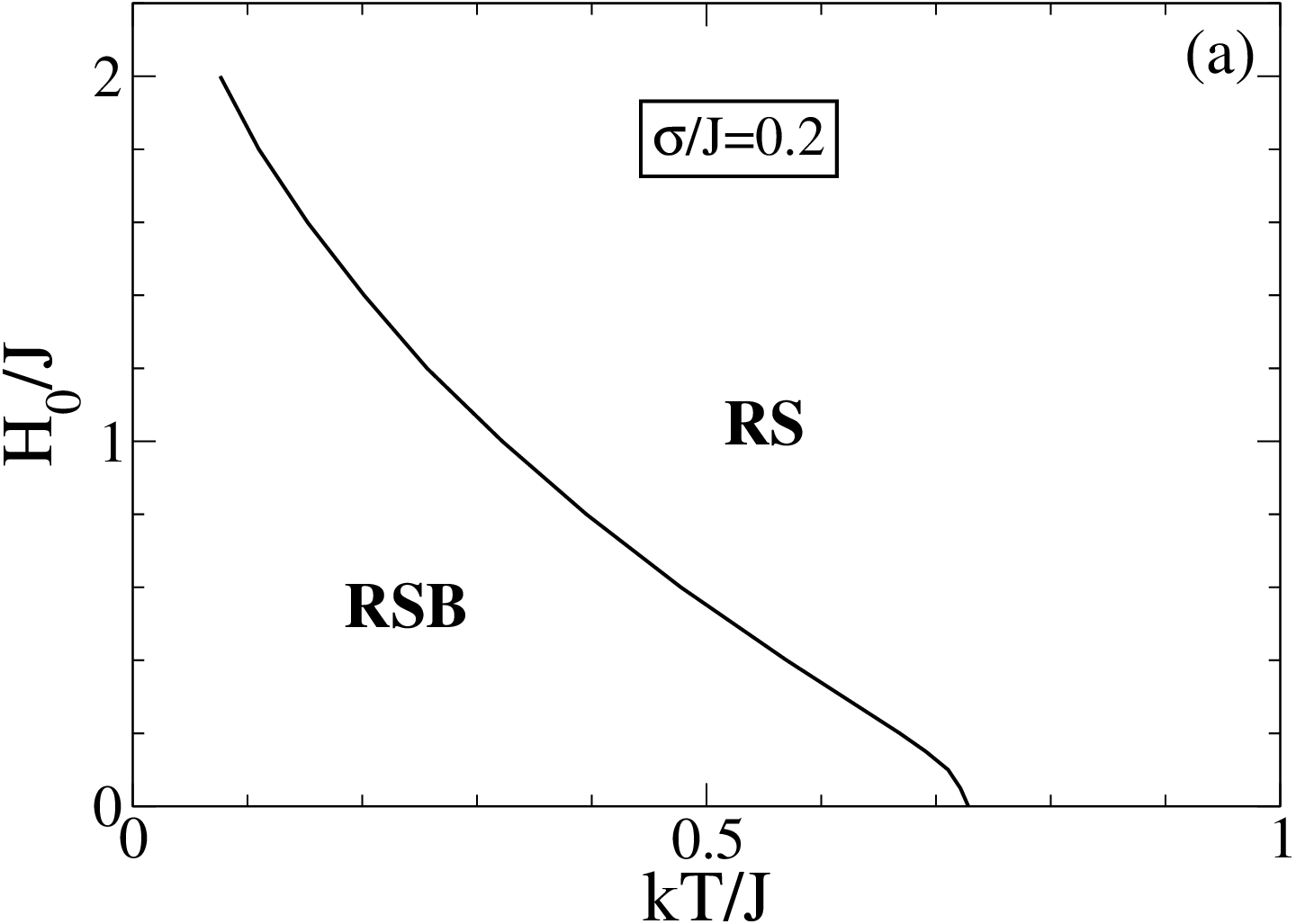}
\hspace{1.5cm}
\includegraphics[width=0.40\columnwidth,angle=0]{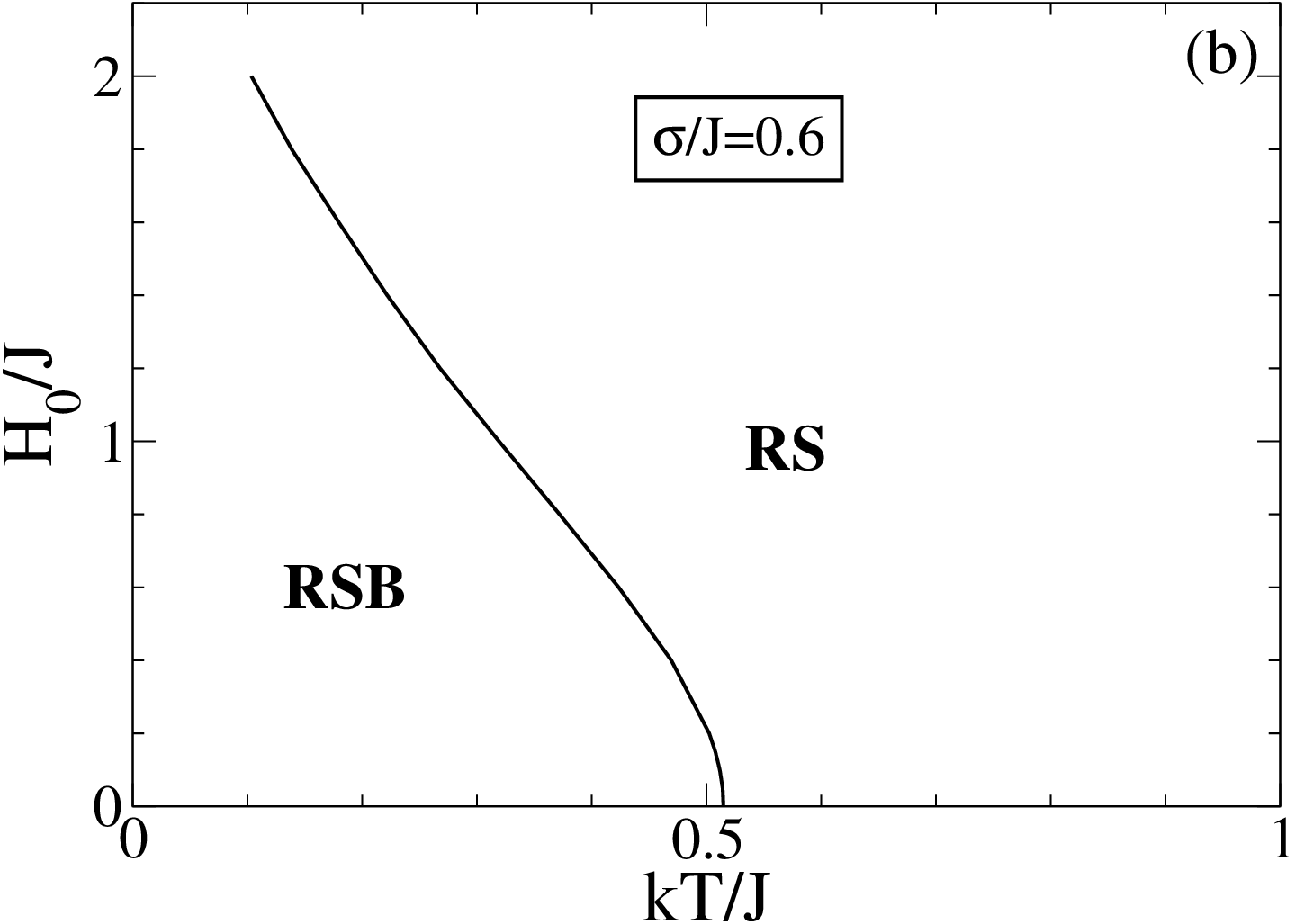}
\end{center}
\protect\caption{Instabilities of the replica-symmetric solution of the  
infinite-range-interaction ISG (cases $J_{0}=0$) in
the presence of a double-Gaussian random field, for two
typical values of distribution widths: (a) $(\sigma/J)=0.2$; (b)
$(\sigma/J)=0.6$. 
In each case the AT line separates a region of RS from the one
characterized by RSB (all variables are scaled in units of $J$).}  
\label{fig8}
\end{figure}

In order to illustrate that the present model is capable of reproducing
qualitatively the phase diagrams of previous works, namely, the Ising spin
glass in the presence of random fields following either a Gaussian
\cite{soares}, or a bimodal \cite{nogueira} probability distribution, in
Fig.~7 we compare typical results obtained for the Ising spin-glass model 
in the presence of a double Gaussian distribution characterized by
$(\sigma/J)=0.4$ 
with those already known for such particular cases. 
In these comparisons, we have chosen qualitatively similar phase diagrams,
mainly taking into 
account the critical frontier that separates the regions with $m=0$ and
$m\neq0$. 
In Fig.~7(a) we exhibit the phase diagram of the present model 
[$(H_{0}/J)=0.8$] together with 
the one of an ISG in the presence of random fields described by a single
Gaussian distribution; both phase diagrams are qualitatively similar to the
one of the standard SK model. 
In Fig.~7(b) we present phase diagrams for the bimodal and double Gaussian
distributions, at their corresponding threshold
values, $H_{0}^{(1)}(\sigma)$. 
Typical situations for the cases of the bimodal and
double Gaussian distributions, where two tricritical points appear
along the critical frontier that separates the regions with $m=0$ and
$m\neq0$, are shown in Fig.~7(c). 
Phase diagrams for the bimodal and double Gaussian
distributions, at their corresponding threshold
values, $H_{0}^{(2)}(\sigma)$, are presented in Fig.~7(d).

Next, we analyze the AT instability for $J_{0}=0$; in this case,
Eq.~(\ref{15}) may be written as  

\begin{equation} \label{36}
\left(\frac{kT}{J}\right)^{2}=\frac{1}{\sqrt{2\pi}}
\int_{-\infty}^{+\infty}dz
e^{-z^{2}/2}{\rm sech}^{4}\beta J\left\{ Gz+\frac{H_{0}}{J} \right\}~,   
\end{equation}

\vskip \baselineskip
\noindent
which corresponds to the same instability found in the case of a
single-Gaussian random field \cite{soares}. In Fig.~8 we exhibit AT lines
for two typical values of distribution widths; in each case the AT line
separates a region of RS from the one characterized by RSB. 
One notices that the region associated with RSB gets reduced for increasing
values of $\sigma$; however, the most interesting
aspect in these lines corresponds to an inflection point, which may be
identified with the one that has been observed in the experimental
equilibrium boundary of the compound ${\rm Fe_{x}Zn_{1-x}F_{2}}$
\cite{montenegro91,soares}. Up to now, this effect was believed to be
explained only through the ISG in the presence of a single-Gaussian random
field, for which the phase diagrams in the  
cases $J_{0}>0$ are much simpler, with all phase transitions being
continuous, typically like those of the SK model.
Herein, we have shown that an inflection point in the AT line may also
occur in the present model, for which one has a wide variety of phase
diagrams in the corresponding case $J_{0}>0$, as exhibited above. 
Therefore, the present model would be appropriate for explaining a similar
effect that may be also   
observed experimentally in diluted antiferromagnets characterized by
first-order phase transitions, like ${\rm Fe_{x}Mg_{1-x}Cl_{2}}$.   

\vskip \baselineskip
\noindent
\section{Conclusions}

\vskip \baselineskip

We have studied an Ising spin-glass model, in the limit of infinite-range
interactions and in the presence of random magnetic fields distributed  
according to a double-Gaussian probability distribution. Such a
distribution contains, as particular limits, both the single-Gaussian and
bimodal probability distributions. By varying the parameters of this
distribution, a rich variety of phase diagrams is obtained, with continuous
and first-order phase transitions, as well as tricritical points.
The condition for the existence of a single finite-temperature tricritical
point at the 
critical frontier Paramagnetic-Ferromagnetic (i.e., in the region of
stability of 
the replica-symmetric solution), characterized by a first-order line at low
temperatures, is derived analytically.   
Besides that, we found an inflection point in the AT line (in the
plane magnetic field versus temperature), which may correspond to the one
observed in the compound ${\rm Fe_{x}Zn_{1-x}F_{2}}$
\cite{montenegro91,soares}. 
This effect, which has already shown up in the Ising spin-glass model in
the presence of a Gaussian random field \cite{soares}, is herein obtained
for a more 
general probability distribution for the magnetic fields. Hence, 
the present model is appropriate for explaining
a similar effect that could be   
observed also in diluted antiferromagnets characterized by first-order
phase transitions, like ${\rm Fe_{x}Mg_{1-x}Cl_{2}}$.   
Therefore, with this random-field distribution, one
may adjust the model to given physical situations, in order to
reproduce a wide diversity of effects that occur in real systems. 

The double-Gaussian probability
distribution, defined above, is suitable for a theoretical
description of random-field systems, being a better candidate for 
such a purpose than the two most commonly used distributions in the 
literature, namely, the bimodal and single-Gaussian distributions, due to
the following reasons:    
(i) In the identifications of the RFIM with diluted
antiferromagnets in the presence of a uniform magnetic field, the local
random fields are expressed in terms of quantities that vary in both sign
and magnitude \cite{fishmanaharony,cardy}.  
This characteristic rules out
the bimodal probability distribution from such a class of physical systems.
The double-Gaussian probability distribution is appropriate for a
description of diluted antiferromagnets for a large range of magnetic
concentrations,  
like in the RFIM, as well as in the ISG regimes.   
(ii) Although the RFIM defined in terms of a single-Gaussian probability
distribution for the fields is physically acceptable, it usually
leads to a continuous phase transition at finite temperatures, either within
mean-field \cite{aharony78,andelman,galambirman}, or standard
short-range-interaction approaches \cite{gofman,swift}. 
Such a system is not able to exhibit
first-order phase transitions and tricritical points, that may occur in some
diluted antiferromagnets \cite{belangerreview}. 
A similar behavior was obtained for an ISG in the presence of random 
magnetic fields following a single-Gaussian probability distribution, where
all phase transitions were found to be continuous \cite{soares}.  The
present model, defined in terms of a double-Gaussian probability
distribution, is expected to be relevant for ${\rm Fe_{x}Mg_{1-x}Cl_{2}}$
(which is known to exhibit a first-order phase transition in the RFIM
regime 
\cite{belangerreview}) with concentrations $x<0.55$, where the ISG behavior
shows up. 

\vskip 2\baselineskip

{\large\bf Acknowledgments}

\vskip \baselineskip
\noindent
We thank Prof. E.~M.~F. Curado for fruitful
conversations. The partial financial supports from
CNPq and Pronex/MCT/FAPERJ (Brazilian agencies) are acknowledged. 

\vskip 2\baselineskip

\end{document}